\documentclass[11pt]{article}
\pdfoutput=1 

\usepackage{jheppub} 

\usepackage[T1]{fontenc} 
\usepackage{multirow}
\usepackage{youngtab}

\makeatletter
\g@addto@macro\bfseries{\boldmath}
\def\NAT@sort{\z@}
\makeatother

\title{\boldmath  Large U(1) charges in F-theory}

\author[1,2]{Nikhil Raghuram}
\author[2]{Washington Taylor}

\affiliation[1]{Department of Physics\\
  Robeson Hall, 0435\\
  Virginia Tech\\
  850 West Campus Drive\\
  Blacksburg, VA 24061, USA}
\affiliation[2]{Center for Theoretical Physics\\
Department of Physics\\
Massachusetts Institute of Technology\\
77 Massachusetts Avenue\\
Cambridge, MA 02139, USA}

\emailAdd{{\tt nikhilr} {\rm at} {\tt vt.edu}}
\emailAdd{{\tt wati} {\rm at} {\tt mit.edu}}

\preprint{MIT-CTP-5053}

\newcommand{\gsu}[0]{\text{SU}}
\newcommand{\gu}[0]{\text{U}}

\DeclareMathOperator{\Tr}{Tr}
\DeclareMathOperator{\tr}{tr}
\newcommand{\chargehyper}[1]{\mathbf{(q=#1)}}
\newcommand{\pmq}[1]{\pm #1}
\newcommand{\vev}[0]{VEV}
\newcommand{\uoneb}[0]{\tilde{b}}
\newcommand{\maxcharge}[0]{\pmq{21}}

\abstract{We show that massless fields with large abelian charges (up
  to at least $q = 21$) can be constructed in 6D F-theory models with
  a U(1) gauge group.  To show this, we explicitly construct F-theory
  Weierstrass models with nonabelian gauge groups that can be broken
  to U(1) theories with a variety of large charges.  Determining the
  maximum abelian charge allowed in such a theory is key to
  eliminating what seems currently to be an infinite swampland of
  apparently consistent U(1) supergravity theories with large
  charges.}

\begin{document} 
\maketitle
\flushbottom

\section{Introduction}

\subsection{Charged matter in string theory and F-theory}

While string theory can produce a vast range of consistent
supergravity theories in four and higher space-time dimensions, there
are nonetheless constraints on what kinds of low-energy theories can
arise from string theory.
These constraints are particularly strong in higher dimensions, and
have recently been explored in 10D \cite{10D-universality}, 8D
\cite{Taylor:2011wt, 8D-anomalies}, and 6D
\cite{6D-universality, KMT-2, SeibergTaylorLattices,Grimm:2015zea, KMRT,
CveticLinU1,  AT-WT-infinite, MonnierMooreParkQuantization, Lee:2018urn}.
 More generally,
the set of low-energy theories that look consistent but cannot be
realized in string theory have been referred to as the ``swampland''
\cite{swamp, Ooguri-Vafa-swamp}.

A particularly interesting question that is relevant in every
dimension is: what kinds of light or massless matter fields can arise
in compactifications of string theory?  A priori, one might think that
a matter field could transform under any representation of a gauge
group $G$ in a consistent low-energy theory of gravity.  This is not
the case, however,  at least in higher dimensions with supersymmetry.
The highest dimension in which matter fields can arise in any
representation other than the adjoint in a supersymmetric theory is
6D.  In this paper, we consider possible charges for massless fields
charged under a gauge group
in a 6D supergravity theory that has only a single U(1) factor, like
the familiar four-dimensional theory of electromagnetism.

In six dimensions, for supersymmetric theories of gravity with
nonabelian gauge groups, there are strong constraints on the possible
matter representations that can arise.  For theories with fewer than 9
tensor multiplets, anomaly cancellation conditions alone restrict the
set of possible nonabelian gauge groups and charged matter fields to a
finite set \cite{kt-finite, KMT-2}.  In six dimensions, F-theory \cite{Vafa-F-theory,
MorrisonVafaI, MorrisonVafaII} (see \cite{Taylor:2011wt,Weigand:2018rez,Cvetic:2018bni} for recent reviews and further background on F-theory) gives the most general class of known
supersymmetric
string vacuum constructions, and recent work has focused on what kinds
of matter representations can be realized for massless fields in 6D
F-theory models.  The simplest F-theory models give only a simple generic set of massless matter fields; for an SU(N) gauge group, these fields are in just the singlet, fundamental, adjoint, and two-index antisymmetric representations. A
few more exotic representations can be constructed in F-theory (see
\cite{KMRT,  Cvetic:2018xaq} for recent work and further
references), but the constraints
both from anomaly cancellation and from F-theory on the allowed
representations are generally quite strong.  For example, it seems
that in F-theory no massless matter field can transform in any
representation of SU(2) of dimension higher than 5.

For abelian charges, however, the story is quite different and less
well understood.  As far as low-energy consistency conditions go,
there seems to be an infinite family of
6D supergravity models with a U(1) gauge group, even in theories
without tensor multiplets, in which  the abelian charges $q$ can be
arbitrarily large \cite{AT-WT-infinite}.  On the other hand, from the
finiteness of the set of elliptic Calabi-Yau threefolds \cite{Gross,
  KMT-2}, it is clear that there is a finite upper bound on the
largest abelian charge $q_{\rm max}$ that can be realized in any
F-theory construction.  This raises the natural
question of what is the largest abelian charge that can arise in a 6D
F-theory vacuum with a single-factor U(1) gauge group.  Little is
known about the answer to this question.  The most well understood F-theory
U(1) models \cite{Morrison-Park} have abelian charges of only $q = 1,
2$. Explicit F-theory models with abelian charge $q = 3$ were first
found in \cite{KleversMayorgaPenaOehlmannPiraguaReuterToric} and constructed more generally in \cite{Nikhil-34},
along with some explicit models with abelian charge $q = 4$.  The
F-theory models with larger abelian charges, however, contain
increasingly complicated singularity structures, and are hard to
analyze analytically.

In this paper, we
use an indirect method to show that F-theory must allow the
construction of 6D supergravity theories with a U(1) gauge group and
massless fields with large abelian charges. Our strategy is to
explicitly construct F-theory models with nonabelian gauge groups
that, according to field theory arguments, can be Higgsed to $\gu(1)$
groups admitting large charges. 
With this
technique, we show that F-theory admits abelian charges as large as
$q=21$.

In Section \ref{sec:strategy}, we describe our strategy in further
detail.  In Section \ref{sec:higgsprocess}, we discuss some general
aspects of Higgsing processes and present a specific Higgsing chain
that is used throughout the rest of the analysis. Section
\ref{sec:anomcancel} reviews the 6D anomaly cancellation conditions
for $\gsu(N)$ and $\gu(1)$ gauge groups, which are then used in
Section \ref{sec:F-theoryanom} to constrain the scope of F-theory
models considered. We then turn to explicit constructions of the
F-theory models with nonabelian gauge groups. Section \ref{sec:p2}
focuses on F-theory models on a $\mathbb{P}^2$ base, which allow us
to demonstrate
that abelian charges $q=1$ through $7$ are realized in
F-theory. In Section \ref{sec:fn}, we discuss F-theory models on
Hirzebruch surfaces
$\mathbb{F}_n$ bases. These models allow us to realize the largest
abelian charges found in this paper. We conclude in Section
\ref{sec:conclusion} by presenting some open questions and directions
for future work.

\subsection{General strategy}
\label{sec:strategy}
Ideally, one would establish that a certain charge can be realized in
F-theory by finding an explicit $\gu(1)$ model admitting the desired
charge. However, constructing F-theory models with large charges is a
challenging enterprise. Weierstrass models admitting
charges\footnote{Note that in 6D a matter hypermultiplet containing a
  field of charge $q > 0$ also contains a field of charge
$-q$.} larger
than $q=\pmq{2}$ involve algebraically complex non-UFD structures. As
one attempts to obtain larger and larger charges, the Weierstrass
models become more and more unwieldy. The currently known F-theory
models with just a $\gu(1)$ gauge group only admit charges $\pmq{1}$
through $\pmq{4}$, and there are few, if any, tractable techniques
available for
systematically constructing models with arbitrarily large charges.

Given these difficulties, we use an indirect approach to determine
that any other specific charges must be realized in F-theory. Our strategy is to
explicitly realize F-theory models with nonabelian gauge groups that
can be Higgsed down to a $\gu(1)$ gauge symmetry admitting large
charges. In particular, we focus on 6D F-theory models having an
$\gsu(N)$ gauge group and at least two adjoint hypermultiplets. As
described in \S\ref{sec:higgsprocess}, such a low-energy $\gsu(N)$
supergravity model can be Higgsed down to a $\gu(1)$ model. If the
$\gsu(N)$ model can be realized in F-theory, it must be therefore possible to
deform it to the corresponding $\gu(1)$ F-theory model. In other
words, constructing the $\gsu(N)$ model in F-theory demonstrates that
the corresponding $\gu(1)$ model must exist in F-theory, even if we
cannot determine the exact deformations necessary to Higgs the
$\gsu(N)$ symmetry. And if field-theoretic considerations
show that
the $\gu(1)$ theory has hypermultiplets with large charges, those
large charges must be realizable in F-theory.

We therefore focus on constructing explicit $\gsu(N)$ Weierstrass
models. The Higgsing process, which can be understood purely from
field-theoretic considerations, then implies that certain $\gu(1)$
charges can be realized. Thus, we can establish that particular
charges occur in F-theory without explicitly constructing $\gu(1)$
F-theory models. Of course, this strategy has some limitations. While
we can show that certain charges occur in F-theory, we cannot prove that certain charges are ruled out in F-theory. As a
result, we will not be able to establish an upper bound on the charges
in F-theory. Nevertheless, this technique demonstrates that the
highest possible charge must be at least $\maxcharge{}$, significantly larger than the charges that have currently been realized in explicit F-theory models. Even if they
cannot rule out certain charges, these $\gsu(N)$ models provide new
information about the possible charge spectra in F-theory.

Note that in this paper when we speak of ``large'' U(1) charges, we
mean relative to the natural unit of charge in the theory.  In most
cases we deal with the natural unit of charge is the greatest common
divisor of the nonzero massless charges, and is generally 1 in the units we
use. We discuss this issue a little further in \S\ref{sec:anomcancel}.

\section{The Higgsing process, anomalies, and F-theory models in 6D}

In this section we go over some basic aspects of  the
Higgsing process, constraints from anomaly cancellation, and F-theory models for 6D supergravity
theories.
In general, 6D supergravity theories have some number $T$ of tensor
multiplets, a gauge group $G$, and hypermultiplet matter fields
transforming in some representation ${\cal R}$ of $G$.  Here,
we focus primarily on theories with zero or one tensor multiplets ($T
= 0, 1$), and gauge groups of the form SU($N$), U(1), or products of
such factors.  In particular, we are interested in starting with a
theory having a gauge group SU($N$) and at least two matter fields in
the adjoint representation, which can be broken by Higgsing processes
down to a theory with a U(1) gauge group and various charged matter representations.

\subsection{The Higgsing process}
\label{sec:higgsprocess}

We begin with a few generalities on Higgsing processes in 6D
theories with ${\cal N} = (1, 0)$ supersymmetry.  While there are some differences, such processes can be
understood in analogy with Higgsing processes in ${\cal N} = 1$ 4D
  gauge theories.
 In the latter context, a process in which a field or fields $\phi_i$
acquire nonzero expectation values and break a gauge group $G$ can be
described either in terms of supersymmetric D-term constraints or
geometric invariant theory. In the former context, the field
expectation values must satisfy the conditions
$\sum_{i}\phi_i^{\dagger}T^{A}\phi_i = 0$, where each generator
$T^{A}$ acts on the fields $\phi_i$ according to the appropriate
representation.  In the context of geometric invariant theory, the
vacua are parameterized by gauge-invariant polynomials in the fields
$\phi_i$ \cite{Luty-Taylor}. From each point of view one can see that
 a gauge group can be broken by Higgsing on a single field in the
adjoint representation; for example, from the D-term point of view
this follows from the fact that $T^{A}$ acts through the adjoint
action, so that the D-term conditions automatically vanish.  On the
other hand, one needs two fields in the fundamental representation to
break SU($N$) through Higgsing, since a single nonzero VEV cannot
solve the D-term constraints for all generators, and cannot be used to
form a gauge-invariant polynomial.

Our
ultimate goal is to break gauge groups, such as $\gsu(N)$, down to
$\gu(1)$ in a way that generates large charges. To accomplish this, we
use a specific $\gsu(N)\rightarrow \gu(1)$ Higgsing process described
in \cite{AT-WT-infinite}. The preserved $\gu(1)$ corresponds to the
$\gsu(N)$ generator
\begin{equation}\text{diag}(1,1,\ldots,1,-N+1), \label{eq:preservedgenerator}
\end{equation}
which is written in the fundamental representation. The details of this Higgsing process are summarized below for convenience. In most instances, we consider the resulting charge spectrum when an $\gsu(N)$ gauge group undergoes this exact Higgsing process. Even when we consider alternative Higgsing processes, the steps outlined below form part of the Higgsing sequence. 

The starting point for this Higgsing process is a 6D supergravity theory with an $\gsu(N)$ gauge symmetry and at least two hypermultiplets in the adjoint representation. Giving a generic \vev{} to one of the adjoint hypermultiplets breaks $\gsu(N)$ to its Cartan subgroup, $\gu(1)^{N-1}$. We can describe how a hypermultiplet is charged under this $\gu(1)^{N-1}$ symmetry using a charge vector $\vec{q}=(q_1,\ldots,q_{N-1})$, where $q_i$ denotes the charge under the $i$th $\gu(1)$. After giving a \vev{} to the first adjoint multiplet, an $\gsu(N)$ representation $\mathbf{R}$ branches to a collection of $\gu(1)^{N-1}$ charge vectors corresponding to the weight vectors of $\mathbf{R}$.

We then want to give \vev{}s to hypermultiplets charged under the
$\gu(1)^{n-1}$ symmetry to break it to a single $\gu(1)$. This is not
possible if one uses the remnant hypermultiplets from only the first
adjoint hypermultiplet. Most of the degrees of freedom in this first
adjoint hypermultiplet are eaten, and those that remain after the
Higgsing are neutral under the $\gu(1)^{N-1}$ symmetry. But all the
degrees of freedom from the second adjoint hypermultiplet still
remain, and many of them are charged under the $\gu(1)^{N-1}$
symmetry. After giving the \vev{} to the first adjoint hypermultiplet, the second adjoint
hypermultiplet branches to $N^2-N$ charged hypermultiplets whose charge
vectors $\vec{q}=(q_1,\ldots,q_{N_1})$ under $\gu(1)^{N-1}$ are the
$\gsu(N)$ root vectors.\footnote{In addition to these $N^2-N$ charged
  hypermultiplets, there are $N-1$ neutral hypermultiplets coming from
  the second adjoint hypermultiplet.} We can
work in the Dynkin basis, where the root vectors for the simple roots
are the rows of the Cartan matrix:
\begin{align}
\alpha_1 =& (2,-1,0,0,\ldots) \notag\\
\alpha_2 =& (-1,2,-1,0,\ldots) \notag\\
\alpha_3 =& (0,-1,2,-1,\ldots) \\ 
&\vdots \notag\\
\alpha_{N-1} =& (0,\ldots,0,-1,2) \notag
\end{align}

We can now give \vev{}s to the charged hypermultiplets whose charge
vectors are the simple roots $\alpha_2$ through
$\alpha_{N-1}$ (along with their negative counterparts to satisfy the
D-term constraints). This breaks the
$\gu(1)^{N-1}$ symmetry down to a single $\gu(1)$ corresponding to the
direction in root space orthogonal to $\alpha_2$ through
$\alpha_{N-1}$. In the Dynkin basis, this direction is given by the
vector $(N-1, N-2, \ldots, 2, 1)$.

This explicit description of the Higgsing processs allows us to calculate the resulting $\gu(1)$ charges. For instance, the fundamental representation of $\gsu(N)$ consists of weights of the form
\begin{equation}
[1,0,0,\ldots], [-1,1,0,0\ldots], [0,-1,1,0\ldots], \ldots [0,\ldots,0,-1,1].
\end{equation}
When one takes the inner product of these weights with $(N-1, N-2, \ldots, 2, 1)$, the highest weight $[1,0,0\ldots]$ leads to charge $N-1$, while the other weights lead to charge $-1$. These charges agree with the diagonal entries in \eqref{eq:preservedgenerator}, at least up to sign and normalization, indicating that we have preserved the desired generator. Hypermultiplets in a representation $\mathbf{R}$ include fields in both $\mathbf{R}$ and $\mathbf{\overline{R}}$, and the hypermultiplets charged under the final $\gu(1)$ include fields with both positive and negative charges. Therefore, a fundamental hypermultiplet branches to $\gu(1)$ hypermultiplets in the following way:
\begin{equation}
{\tiny \yng(1)} \rightarrow \chargehyper{N-1} + (N-1)\times\chargehyper{1}.
\end{equation}
Note that this result can also be derived easily in the fundamental
basis; if the unbroken adjoint field takes the form diag$(1, 1, \ldots,
-N + 1)$, then clearly a field in the fundamental representation breaks
up into $N - 1$ fields of charge $\chargehyper{1}$ and one field of
charge $\chargehyper{N -1}$.
All the calculations here can be carried out in a straightforward
fashion in either basis; the Dynkin basis may be more useful for
generalization to other groups.

Similar calculations show that other $\gsu(N)$ representations branch
as\footnote{Even though these formulas allow one to compute all of the
  resulting spectra in this paper by hand, many of the calculations
of specific spectra quoted later were also performed using LieART \cite{Feger:2012bs} as an
  additional check.}
\begin{align}
\textbf{Adj}\rightarrow& 2(N-1) \times \chargehyper{N} + \left(N-1\right)^2\times\chargehyper{0}\\
{\tiny \yng(1,1)}\rightarrow& \left(N-1\right)\times\chargehyper{N-2}+\frac{(N-1)(N-2)}{2}\times\chargehyper{2}\\
{\tiny\yng(1,1,1)}\rightarrow& \frac{(N-1)(N-2)}{2}\times\chargehyper{N-3} + \frac{(N-1)(N-2)(N-3)}{6}\chargehyper{3}
\end{align}

Table \ref{tab:chargesummary} summarizes the charges coming from
different $\gsu(N)$ representations under this Higgsing
process.\footnote{While we restrict our attention to the
  representations listed in Table \ref{tab:chargesummary}, one could
  consider other representations, namely the symmetric
  representation. Under the Higgsing process that we have described here, the symmetric
  representation would give charges as large as
  $\pmq{(2N-2)}$. However, if we require that there are at least two
  adjoints, the largest $\gsu(N)$ model that we have been able to obtain using
  techniques similar to those in \cite{KMRT} is $\gsu(5)$, at least
  for a $\mathbb{P}^2$ base. Therefore, including symmetric matter
  does not provide an obvious way of obtaining significantly larger
  charges, although it would be interesting to systemically explore
  the charges possible when one includes the symmetric representation;
  we leave this for future work.}
 Already, one can make interesting observations
about the charge spectra. Many of the $\gsu(N)$ gauge groups lead to
massless charged spectra that skip over certain charges. For instance,
consider Higgsing an $\gsu(8)$ model with hypermultiplets in the
representations listed in Table \ref{tab:chargesummary}. The resulting
$\gu(1)$ charge spectrum includes all of the charges from $\pmq{1}$ to
$\pmq{8}$ \emph{except} for charge $\pmq{4}$. This fact might naively
seem to contradict the completeness hypothesis
\cite{Banks:2010zn, HarlowOoguri}, which states that all possible charges must be realized in
the Hilbert space of a theory. However, as discussed in
\cite{AT-WT-infinite}, the charge spectra we consider here involve
only massless states, whereas the completeness hypothesis considers
both massless and massive states. Massless $\gu(1)$ charge spectra
that seem to skip over charges therefore do not directly contradict
the completeness hypothesis. Nevertheless, one might be tempted to
conjecture that F-theory $\gu(1)$ models obey some massless equivalent
of the completeness conjecture in which all charges between $\pmq{1}$
and some large value occur. Examples such as the $\gsu(8)$ model
above, which we explicitly construct in F-theory in \S\ref{sec:fn},
contradict these sort of conjectures.

\begin{table}
\begin{center}
\begin{tabular}{ccccc}
\textbf{Gauge} & \multirow{2}{*}{\emph{Fundamental}}  & \multirow{2}{*}{\emph{Adjoint}} & \emph{Two-Index}   & \emph{Three-Index} \\
\textbf{Group}& & & \emph{Antisymmetric}  & \emph{Antisymmetric}  \\\hline
$\gsu(2)$ & $\pm 1$& $0$, $\pm 2$ & ---  & --- \\
$\gsu(3)$ & $\pm 1$, $\pm 2$ & $0$, $\pm 3$& ---  & ---\\
$\gsu(4)$ & $\pm 1$, $\pm 3$ & $0$, $\pm 4$ & $\pm 2$  & --- \\
$\gsu(5)$ & $\pm 1$, $\pm 4$ & $0$, $\pm 5$ & $\pm 2$, $\pm 3$& ---\\
$\gsu(6)$ & $\pm 1$, $\pm 5$ & $0$, $\pm 6$ &$\pm 2$, $\pm 4$ &  $\pm 3$\\
$\gsu(7)$ & $\pm 1$, $\pm 6$ & $0$, $\pm 7$ & $\pm 2$, $\pm 5$  & $\pm 3$, $\pm 4$\\
$\gsu(8)$ & $\pm 1$, $\pm 7$ & $0$, $\pm 8$ & $\pm 2$, $\pm 6$ & $\pm 3$, $\pm 5$\\
\end{tabular}
\end{center}
\caption{$\gu(1)$ charges realized by Higgsing $\gsu(N)$ according to
  the Higgsing process on two adjoints described in the text. Each entry denotes the charges coming from a
  hypermultiplet in a particular representation of $\gsu(N)$. Note
  that hypermultiplets include fields in a representation $\mathbf{R}$
  and its conjugate $\mathbf{\bar{R}}$, allowing one to obtain both
  positive and negative charges from a single hypermultiplet. Dashes
  indicate representations that either do not occur for a particular
  gauge group or are equivalent to some other representation. Not all
  of the representations listed in this table appear in the F-theory
  models considered later.}
\label{tab:chargesummary}
\end{table}

It is important to note that this Higgsing
process can be seen directly
and explicitly in $\gsu(3)$ and $\gsu(4)$ F-theory
models \cite{Nikhil-34}. In Appendix \ref{sec:expHiggsingSU4}, we give
a specific example in which an a $\gu(1)$ F-theory model with charge
$\pmq{4}$ matter is unHiggsed to an $\gsu(4)$ model admitting two
adjoint hypermultiplets. The explicit realization of this
Higgsing/unHiggsing process provides additional confirmation of the
general arguments presented above. Of course, our ultimate goal is
determine whether larger charges can be realized in F-theory. Table
\ref{tab:chargesummary} already suggests that $\gsu(N)$ models should
lead to charges beyond those currently realized in F-theory $\gu(1)$
models. But before we can establish that certain $\gu(1)$ charges
occur in F-theory, we must show that the corresponding $\gsu(N)$
models can be realized in F-theory. We turn to this issue next.

\subsection{Anomaly cancellation conditions for $\gsu(N)$ and U(1) models}
\label{sec:anomcancel}
Clearly,
knowing the types of $\gsu(N)$ models that can be realized tells us
information about the possible U(1) charges.  We therefore must determine which 6D $\gsu(N)$
supergravity models can be realized in F-theory. In particular, larger
$\gsu(N)$ models allow us to obtain larger $\gu(1)$ charges, so we are
most interested in determining the largest suitable $\gsu(N)$ models
that occur in F-theory with at least two hypermultiplets of adjoint matter. A worthwhile first step is to determine the 6D
$\gsu(N)$ supergravity models that satisfy the anomaly cancellation
conditions. All F-theory constructions should satisfy these
conditions, allowing us to narrow the scope of F-theory models to
investigate. Of course, a model that satisfies the anomaly
cancellation conditions may not have an F-theory
realization.  Nevertheless, the anomaly analysis provides interesting
insights into the $\gsu(N)$ F-theory models and their implications for
the $\gu(1)$ charge spectra.

6D (1,0) supergravity theories have chiral spectra and may therefore
suffer from anomalies. 
These anomalies can be canceled via the
Green-Schwarz mechanism
\cite{GreenSchwarzWest6DAnom, SagnottiGS}, which uses  tree-level diagrams involving
tensors to cancel contributions from chiral fermions. However, the
massless spectrum must satisfy certain conditions for the anomalies to
cancel. Suppose that our theory has one graviton multiplet, $T$
tensor multiplets, $V$ vector multiplets, and $H$
hypermultiplets. Gravitational anomalies are canceled only if
\begin{equation}
H - V + 29 T = 273.
\end{equation}

Gauge anomalies need to be canceled as well. Let us first focus on
cases where the gauge group is $\gsu(N)$ to simplify the anomaly
cancellation conditions. Suppose that there are $x_{R}$ full
hypermultiplets in the $R$ representation of $\gsu(N)$. The gauge
anomaly conditions depend on two vectors, $a$ and $b$, living in a
lattice $\Gamma$ of signature $(1,T)$ with an inner product denoted by
$\cdot$. 
  Gravitational anomaly cancellation imposes the condition that
$a \cdot a = 9-T$.
Gauge and mixed gauge-gravitational anomalies cancel if the
following equations are satisfied:
\begin{align}
-a\cdot b =& \frac{1}{6}\left(\sum_{R}x_R A_R-A_{\text{adj}} \right),\\
0 =& \sum_{R}x_R B_R - B_{\text{adj}},\\
b\cdot b =& \frac{1}{3}\left(\sum_{R} x_R C_{R}- C_{\text{adj}} \right).
\end{align}
Here we have used the
group theory coefficients $A_R$, $B_R$, and $C_R$ defined by the relations
\begin{align}
\Tr_{R}F^2 =& A_{R}\tr F^2 & \Tr_{R}F^4 =& B_{R}\tr F^4 + C_{R}\left(\tr F^2\right)^2,
\end{align}
where $\tr$ represents a trace in the fundamental representation and $\Tr_{R}$ represents a trace in the $R$ representation. 

If the gauge group is the product of $\gsu(N)$ factors, there is an additional anomaly constraint. Suppose we consider two of the $\gsu(N)$ factors, $\gsu(N)_i$ and $\gsu(N)_j$, with corresponding vectors $b_i$ and $b_j$. Let $x_{(R_i,R_j)}$ denote the number of hypermultiplets in the representation $(R_i,R_j)$ of $\gsu(N)_i\times\gsu(N)_j$. Then, the additional anomaly constraint takes the form
\begin{equation}
  b_i\cdot b_j =\sum_{(R_i,R_j)}x_{(R_i,R_j)}A_{R_i}A_{R_j}.
  \label{eq:additional}
\end{equation}

For theories with a U(1) gauge group, the anomaly conditions take a
similar but simpler form
\cite{Erler6DAnom,ParkTaylorAbelian},
\begin{align}
a \cdot \tilde{b} &= -\frac{1}{6} \sum_{i}  q_i^2\,,
    \label{eq:U1ACsqr} \\
\tilde{b} \cdot \tilde{b} &= \frac{1}{3} \sum_{i} 
q_i^4\,.
    \label{eq:U1ACquar}
\end{align}
Here $q_i$ is the U(1) charge of the $i$th  charged
multiplet and $\tilde{b}$ is again a vector in the lattice $\Gamma$.
When there are multiple U(1) factors, or abelian and nonabelian
factors there are further conditions analogous to
(\ref{eq:additional}), but we will not need those here.

Note that for every spectrum that satisfies the abelian anomaly
equations, there is an infinite family of solutions, which can be
achieved by multiplying all charges by $n$ and  multiplying the
anomaly coefficient $\tilde{b}$ by $n^2$.  While these may seem to be
equivalent theories, under a simple rescaling of the charge, the different value of $\tilde{b}$ in the anomaly
lattice distinguishes the theories.
This is related to the fact that in some F-theory models determining the charge unit can be
subtle.  For example, as discussed in \cite{AT-WT-infinite}, there are
two distinct F-theory models with no tensor multiplets that have 108
charges $q = \pm 1$ and $q = \pm 2$ respectively.  The values of
$\tilde{b}$ differ between these theories by a factor of 4.
This can be seen in F-theory from the fact that an unHiggsing of the
U(1) model gives a nonabelian theory with gauge group SO(3) instead of SU(2).

\subsection{6D F-theory models}
\label{sec:F-theoryanom}
While the anomaly cancellation equations are a low-energy condition, the $a$ and $b$ parameters have a geometric interpretation in F-theory. $a$ can be viewed as the canonical class $K_B$ of the base of the F-theory model's elliptic fibration. $b$, meanwhile, can be viewed as the homology class of the divisor on which the $\gsu(N)$ gauge group is tuned. The inner product $\cdot$ then represents the intersection product between homology classes. In fact, one can solve the gauge and mixed anomaly conditions solely in terms of properties of the gauge divisor, such as its self-intersection $n=b\cdot b$ and its arithmetic genus
\begin{equation}
g = 1+\frac{1}{2}b\cdot\left(a+b\right) \label{eq:genusformula}
\end{equation} 
In other words, the charged spectrum can be determined
without specifying the base. Here, we restrict our attention to the
fundamental, the adjoint, the two-index antisymmetric, and three-index
antisymmetric representations. The resulting charged matter spectra
for $\gsu(2)$ through $\gsu(9)$ are given in Table
\ref{tab:chargedmattermult} \cite{Bershadsky:1996nh, Johnson:2016qar}.

\begin{table}
\begin{center}
\begin{tabular}{ccccc}
\textbf{Gauge} & \multirow{2}{*}{\emph{Fundamental}} &
\multirow{2}{*}{\emph{Adjoint}} & \emph{Two-Index} & \emph{Three-Index}
\\ \textbf{Group}& & & \emph{Antisymmetric} & \emph{Antisymmetric}
\\\hline $\gsu(2)$ & $16-16g+6n$ & $g$ & --- & --- \\ $\gsu(3)$ &
$18-18g+6n$ & $g$ & ---& --- \\ $\gsu(4)$ & $16-16g+4n$ & $g$ &
$2-2g+n$ & ---\\ $\gsu(5)$ & $16-16g+3n$ & $g$ & $2-2g+n$ &
---\\ $\gsu(6)$ & $16-16g+2n+r$ & $g$ & $2-2g+n-r$ & $\frac{1}{2}r$
\\ $\gsu(7)$ & $16-16g+n+5r$ & $g$ & $2-2g+n-3r$ & $r$ \\ $\gsu(8)$ &
$16-16g+9r$ & $g$ & $2-2g+n-4r$ & $r$ \\ $\gsu(9)$ & $16-16g-n+14r$ &
$g$ & $2-2g+n-5r$ & $r$
\end{tabular}
\end{center}
\caption{Charged matter multiplicities for $\gsu(N)$ gauge groups. The
  multiplicities are given in terms of the arithmetic genus $g$ and
  the self-intersection $n=b\cdot b$ of the $\gsu(N)$ divisor. Entries
  with a dash indicate representations that are not relevant for the
  gauge group in question. Note that $r$ is a free
  integer.}
\label{tab:chargedmattermult}
\end{table}

The gravitational anomaly condition, however, requires more global
information. The number of tensor multiplets and neutral
hypermultiplets depends on the specific base chosen. Moreover, some
bases have non-Higgsable clusters
\cite{MorrisonTaylorClusters} which contribute to the number of
vector multiplets. Thus, even if there is no matter charged under both
the $\gsu(N)$ group and the non-Higgsable gauge group, the
non-Higgsable cluster can affect the gravitational anomaly
condition. Therefore, to definitively determine which $\gsu(N)$ models
that come from F-theory
are consistent with anomalies, we need to consider specific bases.

\subsubsection{F-theory models with $T = 0$
  (compactifications on $\mathbb{P}^2$)}
We start by considering $\gsu(N)$ F-theory models on
$\mathbb{P}^2$. Models on $\mathbb{P}^2$ have zero tensor multiplets,
and thus the vectors $a$ and $b$ live in a one-dimensional
lattice. Alternatively, one can say that the basis of homology classes
of $\mathbb{P}^2$ consists of a single element $H$ with
self-intersection number 1. There are no genus-two algebraic curves on
$\mathbb{P}^2$, but quartic curves on $\mathbb{P}^2$ have genus
$g=3$. Thus, if we are willing to have extra adjoint hypermultiplets,
we can construct appropriate $\gsu(N)$ models on $\mathbb{P}^2$.

For quartic curves, $g=3$ and $n=16$. Suppose we assume that there are
no three-index antisymmetric hypermultiplets. The three-index
antisymmetric representation is somewhat exotic from an F-theory
perspective, as the corresponding models are more challenging to
construct and involve additional fine-tuning of the Weierstrass
coefficients.  For instance, $\gsu(N)$ models constructed using Tate's
algorithm \cite{Bershadsky:1996nh,Katz:2011qp} typically admit only
the fundamental, two-index antisymmetric, and adjoint
representations. Restricting our attention to these representations is
therefore a natural first step in the analysis. Under this assumption,
Table \ref{tab:chargedmattermult} suggests that for $N$ larger than 6,
the $\gsu(N)$ model has a negative number of fundamental
hypermultiplets. And for higher degree curves, the number of
fundamental hypermultiplets becomes negative for even smaller values
of $N$. This result would naively suggest that $\gsu(6)$ is the
largest consistent $\gsu(N)$ group on $\mathbb{P}^2$ admitting at
least two adjoint hypermultiplets.

But if we relax the assumption that there are no three-index
antisymmetric multiplets, one can obtain higher values of $N$. For
instance, there is an anomaly-free $\gsu(7)$ model on $\mathbb{P}^2$
with three $\mathbf{48}$ hypermultiplets, four $\mathbf{35}$
multiplets, four $\mathbf{7}$ hypermultiplets, and nine
singlets. However, obtaining $\gsu(8)$ groups and beyond on
$\mathbb{P}^2$ with a sufficient number of adjoints appears difficult
and likely impossible: even when we consider all
four of the representations mentioned, one cannot obtain a suitable
$\gsu(8)$ model or beyond without having a negative number of
fundamental or two-index antisymmetric multiplets.\footnote{Even including the two-index
symmetric representation does not make such a model
possible.}
 To obtain higher
$\gsu(N)$, we must consider bases other than $\mathbb{P}^2$.

\subsubsection{F-theory models with $T = 1$ (compactifications on
  $\mathbb{F}_n$)}
\label{sec:fnanomalies}
Models on $\mathbb{F}_n$ have one tensor multiplet, and $a$ and $b$ live on a two-dimensional lattice. The basis for the homology classes consists of two elements, $S$ and $F$, with
\begin{align}
S\cdot S =& -n & S\cdot F =& 1 & F\cdot F =& 0. 
\end{align}
Additionally, we define a homology class $\tilde{S}\equiv S+nF$, with
\begin{align}
\tilde{S}\cdot \tilde{S}&= n & \tilde{S}\cdot S &= 0 & \tilde{S}\cdot F &= 1.
\end{align}
The canonical class is $K_B=-2S-(n+2)F$.

Unlike $\mathbb{P}^2$, at least some of the $\mathbb{F}_n$ have
algebraic curves of genus two. Our analysis will rely in particular on
smooth curves of class $2\tilde{S}$ on $\mathbb{F}_3$, which have
self-intersection $n=12$. As can be verified with
\eqref{eq:genusformula}, such curves have genus $g=2$, and $\gsu(N)$
groups tuned on these curves admit two adjoint hypermultiplets.
Note that the curve $S$ with self-intersection $S\cdot S= -3$ on
$\mathbb{F}_3$ gives a non-Higgsable cluster: an $\gsu_3$ gauge algebra with
no matter.  Since $S\cdot \tilde{S}= 0$, there is no jointly charged
matter, and this non-Higgsable cluster plays no role in the model
except that it contributes an additional 8 vector multiplets to $V$,
which increases the number of matter hypermultiplets available and is
relevant in some extreme cases as we encounter below.

If we assume there are no hypermultiplets of three-index antisymmetric matter, Table \ref{tab:chargedmattermult} suggests that the largest $\gsu(N)$ we can tune on $2\tilde{S}$ on $\mathbb{F}_3$ is $\gsu(6)$. Beyond this, the number of fundamental hypermultiplets would become negative. However, if we include three-index antisymmetric matter, the anomaly cancellation conditions allow for groups as large as $\gsu(9)$ on $2\tilde{S}$. The $\gsu(9)$ model, with a charged spectrum of
\begin{equation}
2\times\mathbf{80} + 2\times\mathbf{84},
\end{equation}
will not be discussed much here for a few different reasons. First,
the $\mathbf{84}$ representation is difficult (perhaps impossible) to obtain in
F-theory; if it can be realized, this representation would likely
involve complicated mechanisms that may not be visible in the
Weierstrass model
\cite{AndersonGrayRaghuramTaylorMiT,KMRT,Cvetic:2018xaq}. Moreover,
this $\gsu(9)$ model would not give interesting charges under the
Higgsing process of \S\ref{sec:higgsprocess}. Even though the
resulting spectrum naively includes charge $\pmq{9}$ matter, the
resulting charges are all multiples of 3, and the true maximum charge
of the resulting spectrum, in the natural charge units, is
$\pmq{3}$. But the $\gsu(8)$ model on $2\tilde{S}$
can be cleanly realized in F-theory, as discussed further in
\S\ref{sec:fn}.

\section{Explicit F-theory models on $\mathbb{P}^2$ (charges $q = 1$
  through 7)}
\label{sec:p2}
So far, we have shown that certain $\gsu(N)$ spectra satisfy the anomaly cancellation conditions and can be Higgsed down to $\gu(1)$. But a model that satisfies the anomaly cancellation conditions may not necessarily be realized in F-theory. Therefore, we now turn to explicit F-theory constructions of $\gsu(N)$ models. This section focuses on F-theory models on a $\mathbb{P}^2$ base, which have no tensor multiplets. In \S\ref{sec:fn}, we discuss F-theory models with $\mathbb{F}_n$ bases, which admit one tensor multiplet.

\subsection{SU(5) and SU(6) (charges 1 through 6)}
As noted earlier, the fundamental, two-index antisymmetric, and adjoint
representations of $\gsu(N)$ are relatively easy to realize in
F-theory. The anomaly analysis in \S\ref{sec:anomcancel} suggests
that, at least for our purposes, the largest consistent $\gsu(N)$
groups admitting only these representations are $\gsu(5)$ and
$\gsu(6)$ in theories with no tensor multiplets. According to Table \ref{tab:chargesummary}, Higgsing these
$\gsu(N)$ models leads to charges $\pmq{1}$ through $\pmq{6}$. Thus,
by explicitly constructing the appropriate $\gsu(5)$ and $\gsu(6)$
F-theory models, we can demonstrate that charges $\pmq{1}$ through
$\pmq{6}$ can be realized in F-theory.

We construct these F-theory models over a $\mathbb{P}^2$ base. In order to obtain the two adjoint hypermultiplets necessary for the Higgsing process, we must tune the $\gsu(N)$ symmetries on a curve $\sigma=0$ of genus $g\geq 2$. As mentioned previously, there are no algebraic curves of genus 2 on $\mathbb{P}^2$, but quartic curves on $\mathbb{P}^2$ have genus 3. We therefore let $\sigma$ be a smooth quartic curve i.e. a curve with homology class $4H$. The resulting $\gsu(N)$ models have three adjoint hypermultiplets, one more than necessary to Higgs the gauge group down to $\gu(1)$.

Fortunately, there is already a known recipe to construct $\gsu(N)$
models with only the three representations mentioned above. 
The simplest construction of a model with gauge group $\gsu(N)$
proceeds by tuning the coefficients $a_i$ in the ``Tate form'' $y^2 +
a_1 yx + a_3y = x^3 + a_2x^2 + a_4x + a_6$ in a way that automatically
guarantees the appropriate Kodaira singularity type for $\gsu(N)$
\cite{MorrisonVafaII, Bershadsky:1996nh, Katz:2011qp}; the models
constructed in this way have precisely the three representations we
want.  A more general approach to tuning Weierstrass models with
$\gsu(N)$ gauge groups directly was developed in
\cite{Morrison:2011mb}; because we will be interested in models with
other representations we follow that approach here.
The
expressions are different for even and odd $N$, so let us focus on the
$\gsu(5)$ model first. According to the formulas in
\cite{Morrison:2011mb}, the $\gsu(5)$ Weierstrass model
is\footnote{The original expressions include a $g_5\sigma^5$ term in
  the $g$ for the Weierstrass model. However, $g_5$ would be
  ineffective for $[\sigma]=4H$. To address this issue, we simply set
  $g_5$ to zero, which does not cause any problems in the Weierstrass
  model.}
\begin{multline}
y^2 = x^3 + \left(-\frac{1}{3}\Phi^2 + \frac{1}{2}\phi_0\psi_2\sigma^2+f_{3}\sigma^{3}\right)x \\
+ \left(\frac{2}{27}\Phi^3 - \frac{1}{6}\Phi\phi_0\psi_2\sigma^2 - \frac{1}{3}\Phi f_{3}\sigma^{3}+\frac{1}{4}\psi_2^2 \sigma^{4}\right), \label{eq:weiersu5tate}
\end{multline}
where $\Phi$ is given by
\begin{equation}
\Phi = \frac{1}{4}\phi_0^2 + \phi_1 \sigma.
\end{equation}
The homology classes for the various parameters are
\begin{align}
[\sigma] =& 4 H & [\Phi] =& -2K_B = 6H \notag\\
[\phi_0]=& -K_B = 3H & [\psi_2] =& -3K_B - 2[\sigma] = H\\
[f_3] =& -4K_B - 3[\sigma] = 0H & [\phi_1]=& -2K_B - [\sigma] = 2H\\.
\end{align}
The discriminant meanwhile is given by
\begin{equation}
\Delta \equiv 4f^3+27g^2 = \frac{1}{16}\sigma^5\Big[\phi_0^4\psi_2\left(\phi_1 \psi_2-\phi_0 f_3\right)+\mathcal{O}(\sigma)\Big]
\end{equation}
$\Delta$ is proportional to $\sigma^5$, while $f$ and $g$ are not
proportional to $\sigma$. Moreover, the split condition
\cite{Tate,Bershadsky:1996nh,Katz:2011qp} is satisfied, as
$\Phi\big|_{\sigma=0}$ is a perfect square. The Kodaira classification
\cite{KodairaII,KodairaIII,MorrisonVafaII,Bershadsky:1996nh,GrassiMorrisonAnomalies}
therefore indicates
that we have tuned an $I_5$ singularity on $\sigma=0$,
signaling the expected presence of an $\gsu(5)$ gauge group. Additionally, the
only other component of the discriminant is an $I_1$ locus, suggesting
that
there are no other nonabelian gauge factors.

To verify the $\gsu(5)$ model's matter spectrum, we first note
that, because $\sigma$ is a smooth curve of genus 3, there are three
adjoint ($\mathbf{24}$) hypermultiplets. The remaining charged
hypermultiplets are localized at codimension-two loci in the
$\mathbb{P}^2$ base with enhanced fiber singularities. Enhancements
occur at $\phi_0=\sigma=0$ and $\psi_2(\phi_1 \psi_2-\phi_0
f_3)=\sigma = 0$. At $\phi_0=\sigma=0$, the singularity type enhances
from $I_5$ to $I^*_1$, indicating that the $[\phi_0]\cdot[\sigma]=12$
points where $\phi_0=0$ and $\sigma=0$ intersect support two-index
antisymmetric ($\mathbf{10}$) multiplets. Finally, at the
$\psi_2(\phi_1 \psi_2-\phi_0 f_3)=\sigma = 0$ loci, the singularity
type enhances to $I_6$. Therefore, the 16 $\psi_2(\phi_1 \psi_2-\phi_0
f_3)=\sigma = 0$ points support fundamental ($\mathbf{5}$)
multiplets. In summary, the charged spectrum for the $\gsu(5)$ model
is
\begin{equation}
3\times\mathbf{24} + 12\times \mathbf{10} + 16\times \mathbf{5},
\end{equation}
in line with the expectations from the anomaly cancellation conditions. 

The Higgsing procedure outlined in \S\ref{sec:higgsprocess} leads to a charged $\gu(1)$ spectrum of
\begin{equation}
16\times\chargehyper{5} + 16\times\chargehyper{4} + 48\times\chargehyper{3}+72\times\chargehyper{2}+64\times\chargehyper{1}.
\end{equation}
This spectrum satisfies the $\gu(1)$ anomaly cancellation conditions
with $\uoneb{} = 5\times 4\times [\sigma]$, as expected
from the analysis in \cite{AT-WT-infinite}, where it was shown that
this kind of Higgsing of an $\gsu(N)$ model gives a U(1) model with
anomaly coefficient $\tilde{b} = N (N -1) [\sigma]$
. We have
therefore explicitly constructed an $\gsu(5)$ model in F-theory that
can be Higgsed down to a $\gu(1)$ model with charges $\pm{1}$ through
$\pm{5}$. This demonstrates that charges $\pm{1}$ through $\pm{5}$ can
be realized in F-theory.

Now let us turn to the $\gsu(6)$ theory, which allows us to demonstrate that charge $\pmq{6}$ matter can be realized in F-theory. There are in fact two ways to obtain an $\gsu(6)$ model. The first approach is to set $\psi_2$ to $0$ in \eqref{eq:weiersu5tate}, giving 
\begin{equation}
y^2 = x^3 + \left(-\frac{1}{3}\Phi^2 + f_3\sigma^3\right)x+\left(\frac{2}{27}\Phi^3-\frac{1}{3}\Phi f_3\sigma^3\right).\label{eq:weiersu6tate}
\end{equation}
This Weierstrass model
again corresponds to the Tate form, and also
 matches that derived from the expressions in \cite{Morrison:2011mb}.\footnote{Again, a $g_6\sigma^6$ term has been dropped because $g_6$ would be ineffective.} The discriminant is now
\begin{equation}
\Delta = -\frac{1}{16}f_3^2\sigma^6\left[\left(\phi_0^2 +4\phi_1\sigma\right)^2-64 f_3\sigma^3\right].
\end{equation}
The $\sigma^6$ factor indicates the expected presence of an $\gsu(6)$ gauge symmetry on $\sigma=0$. Meanwhile, $[f_3]$ is $0H$, so the $f_3^2$ factor does not represent an additional nonabelian gauge group. Thus, the gauge group is simply $\gsu(6)$.

The matter content analysis resembles that for $\gsu(5)$. Since $\sigma=0$ is still a genus 3 curve, there are three adjoint ($\mathbf{35}$) hypermultiplets. And the $\phi_0 = \sigma=0$ loci still contribute twelve two-index antisymmetric ($\mathbf{15}$) multiplets. However, there are no codimension-two loci where the singularity type enhances from $I_6$ to $I_7$, indicating that there are no fundamental $(\mathbf{6})$ hypermultiplets. The charged matter spectrum is therefore
\begin{equation}
3\times \mathbf{35} + 12\times\mathbf{15},
\end{equation}
in agreement with the anomaly cancellation conditions. The corresponding $\gu(1)$ spectrum is
\begin{equation}
20\times\chargehyper{6} + 60\times\chargehyper{4}+120\times\chargehyper{2}.
\end{equation}
This spectrum hints that $\pmq{6}$ matter can be realized in F-theory, although one might argue that this model truly contains $\pmq{1}$, $\pmq{2}$, and $\pmq{3}$ matter because of the common factor between the charges.\footnote{In later examples, charge $\pmq{6}$ matter appears in spectra without a common factor, more rigorously establishing that charge $\pmq{6}$ matter can be realized in F-theory.} As expected, the spectrum satisfies the $\gu(1)$ anomaly conditions with $\uoneb=6\times 5\times[\sigma]$.

The $\gsu(5)$ and $\gsu(6)$ examples considered so far demonstrate
that matter with charges $\pmq{1}$ through $\pmq{5}$, and possibly
$\pmq{6}$, can be realized in F-theory. At this point, there seems to
be an obstruction to tuning larger $\gsu(N)$ gauge groups. All the
ways of enhancing the $\gsu(6)$ singularity of \eqref{eq:weiersu6tate}
force $(f,g,\Delta)$ to simultaneously vanish on $\sigma=0$,
indicating that $\sigma=0$ would no longer support an $\gsu(N)$
symmetry. This observation is in line with 
the expectations from anomaly
cancellation: for $\mathbb{P}^2$ models with only fundamental,
adjoint, and two-index antisymmetric matter, the largest possible
consistent gauge group is $\gsu(6)$. 
$\gsu(6)$ is also the largest $\gsu(N)$ that can be tuned on a quartic
in $\mathbb{P}^2$ using the Tate tuning approach.
Naively, this might suggest that
our approach can at best demonstrate that charges $\pmq{1}$ through
$\pmq{6}$ occur in F-theory. But these results depend on the
artificial assumption that we consider only the fundamental, adjoint,
and two-index antisymmetric representations. While the constructions
become more complicated, there are still consistent F-theory models
with the three-index antisymmetric representation, allowing us to show
that charges larger than $\pmq{6}$ can be realized in F-theory.

\subsection{SU(6) and SU(7) with three-index antisymmetric matter (charges 6 and 7)}
Let us now consider models admitting the three-index antisymmetric representation. The anomaly conditions suggest that, when one includes three-index antisymmetric matter, an $\gsu(N)$ gauge symmetry on a quartic on $\mathbb{P}^2$ can be as large as $\gsu(7)$. According to Table \ref{tab:chargesummary}, the resulting $\gu(1)$ symmetry would support charge $\pmq{7}$ matter. Thus, if we can explicitly construct this $\gsu(7)$ model in F-theory, we know that charge $\pmq{7}$ matter can be realized in F-theory.

To actually find this $\gsu(7)$ F-theory model, we must consider the second method for obtaining an $\gsu(6)$ Weierstrass model from \eqref{eq:weiersu5tate}. Instead of setting $\psi_2$ to zero, we let
\begin{align}
\phi_1 =& f_3 \beta & \phi_0 = \beta \psi_2,
\end{align}
where $[\beta]= 2H$.
The Weierstrass model is now 
\begin{multline}
y^2 = x^3 + \left(-\frac{1}{3}\Phi^2 + \frac{1}{2}\beta\psi_2^2\sigma^2+f_{3}\sigma^{3}\right)x \\
+ \left(\frac{2}{27}\Phi^3 - \frac{1}{6}\Phi\beta\psi_2^2\sigma^2 - \frac{1}{3}\Phi f_{3}\sigma^{3}+\frac{1}{4}\psi_2^2 \sigma^{4}\right) \label{eq:weiersu63as}
\end{multline}
with
\begin{equation}
\Phi = \beta\left(\frac{1}{4}\beta \psi_2^2+f_3\sigma\right),
\end{equation}
and the discriminant is
\begin{equation}
\Delta = -\frac{1}{16}\sigma^6\left[\beta^3\psi_2^4\left(\psi_2^2+f_3^2\beta\right)+\mathcal{O}(\sigma)\right].
\end{equation}
The $\sigma^6$ factor indicates that we have tuned an $\gsu(6)$ symmetry on $\sigma=0$, and since $\sigma=0$ has genus $g=3$, there are three adjoint ($\mathbf{35}$) hypermultiplets in the spectrum. At $\psi_2=\sigma=0$, the singularity type enhances from $I_6$ to $I_2^*$, so these four points contribute four hypermultiplets of two-index antisymmetric ($\mathbf{15}$) matter. The eight $\psi_2^2+ f_3 \beta=\sigma=0$ points, where the singularity types enhances from $I_6$ to $I_7$, contribute eight hypermultiplets of fundamental ($\mathbf{6}$) matter. But there is a third codimension-two locus, $\beta=\sigma=0$, where the singularity type enhances from $I_6$ to $IV^*$, a behavior not seen in the previous models. These eight points contribute eight half-hypermultiplets of three-index antisymmetric ($\mathbf{20}$) matter. In summary, the total charged spectrum is
\begin{equation}
3\times\mathbf{35}+8\times\frac{1}{2}\mathbf{20}+4\times\mathbf{15}+8\times\mathbf{6}.
\end{equation}
The resulting $\gu(1)$ spectrum would be
\begin{equation}
20\times\chargehyper{6} + 8\times\chargehyper{5} + 20\times\chargehyper{4} + 80\times\chargehyper{3} + 40\times\chargehyper{2}+40\times\chargehyper{1}.
\end{equation}
As expected, this $\gu(1)$ spectrum satisfies the anomaly conditions
with $\uoneb = 6\times 5\times[\sigma]$. With this explicit $\gsu(6)$
model, we have unambiguously shown that charge $\pmq{1}$ through
$\pmq{6}$ can be realized in F-theory. Importantly, the greatest
common factor of the charges is 1, indicating that the $\gu(1)$ model
would genuinely have charge $\pmq{6}$ matter.

We can then derive an $\gsu(7)$ Weierstrass model by letting
\begin{align}
\psi_2 =& f_3 \delta & \beta = -\delta^2,
\end{align}
where $[\delta]=H$. 
The Weierstrass model is now 
\begin{multline}
y^2 = x^3 + \left(-\frac{1}{3}\Phi^2 - \frac{1}{2}f_3^2\delta^4\sigma^2+f_{3}\sigma^{3}\right)x \\
+ \left(\frac{2}{27}\Phi^3 + \frac{1}{6}\Phi f_3^2\delta^4\sigma^2 - \frac{1}{3}\Phi f_{3}\sigma^{3}+\frac{1}{4}f_3^2\delta^2\sigma^{4}\right) \label{eq:weiersu73as}
\end{multline}
with
\begin{equation}
\Phi = f_3\delta^2\left(\frac{1}{4}f_3\delta^4-\sigma\right),
\end{equation}
and the discriminant is
\begin{equation}
\Delta=\frac{1}{16}f_3^3\sigma^7\left[2 f_3^2 \delta^8 - 13 f_3 \delta^4\sigma + 64 \sigma^2\right].
\end{equation}
The $\sigma^7$ factor indicates that the gauge group is $\gsu(7)$, while the $f_3^3$ does not signal the appearance of an extra gauge factor since $[f_3]=0H$. Again, there are three adjoint ($\mathbf{48}$) hypermultiplets because $\sigma=0$ is a genus-3 curve. The only codimension-two singularities occur at $\delta=\sigma=0$, where the $I_7$ singularity type enhances to $III^*$. Each $\delta=\sigma=0$ point therefore contributes a three-index antisymetric ($\mathbf{35}$) hypermultiplet and a fundamental ($\mathbf{7}$) hypermultiplet. The charged matter spectrum is therefore
\begin{equation}
3\times \mathbf{48} + 4\times \mathbf{35}+4\times\mathbf{7},
\end{equation}
in line with the expectations from the anomaly conditions.The corresponding $\gu(1)$ charge spectrum would be
\begin{equation}
24\times\chargehyper{7}+4\times\chargehyper{6} + 60\times\chargehyper{4}+80\times\chargehyper{3}+ 24\times\chargehyper{1},
\end{equation}
which, as expected, satisfies the $\gu(1)$ anomaly conditions with $\uoneb=7\times6\times[\sigma]$. The explicit $\gsu(7)$ F-theory construction therefore shows that charge $\pmq{7}$ matter can be realized in F-theory.

For the four types of representations considered so far, $\gsu(7)$ is
the largest $\gsu(N)$ gauge group that can be tuned on a quartic on
$\mathbb{P}^2$. The anomaly conditions suggest that models with $N>7$
  would have a negative number of two-index antisymmetric multiplets and
  would therefore be inconsistent. However, by again expanding the
  scope of constructions considered, we can obtain charges larger than
  $\pmq{7}$. In particular, we then focus on models with
  $\mathbb{F}_n$ bases, for which anomaly cancellation suggests one
  can obtain a satisfactory $\gsu(8)$ group
with two adjoint matter hypermultiplets.

\section{Explicit F-theory models on $\mathbb{F}_n$
(charges up to $q = 21$)}
\label{sec:fn}
\subsection{\gsu(8) with three-index antisymmetric matter (charge 8) }
So far, we have considered curves on $\mathbb{P}^2$ of genus three, which give us one more adjoint hypermultiplet than needed for the Higgsing process. In principle, we require only a genus-two curve to perform the Higgsing. While there are no algebraic curves of genus two on $\mathbb{P}^2$, there are algebraic genus-two curves on some of the $\mathbb{F}_n$, as mentioned in \S\ref{sec:fnanomalies}. For example, a curve $\sigma=0$ of homology class $2\tilde{S}=2S+6F$ on $\mathbb{F}_3$ has genus
\begin{equation}
g= 1 + \frac{1}{2}[\sigma]\cdot\left(K_B+[\sigma]\right) = 1+\tilde{S}\cdot F = 2.
\end{equation}
The smaller genus allows us to obtain larger $\gsu(N)$ groups on $2\tilde{S}$, which in turn suggest higher charges should exist in F-theory.

In particular, we can tune $\gsu(8)$ on $2\tilde{S}$, implying that
charge $\pmq{8}$ matter can occur in F-theory. To construct the
explicit model, we introduce a coordinate $u$ of homology class $\tilde{S}$ and a
coordinate $v$ of homology class $S$. We start with an $\gsu(6)$
Weierstrass model, which can be constructed by using the formulas in
\cite{Morrison:2011mb} and accounting for the fact that certain
parameters are reducible, which leads in particular to the explicit
appearance of powers of $v$ in $f$ and $g$:
\begin{equation}
f = -\frac{1}{48}v^2\Big[\alpha^4\beta^4 v^2 + 8\alpha^2 \beta^3 \nu v^2 \sigma + 8\beta\left(2\beta \nu^2 v^2 +\alpha^2 \phi_2\right)\sigma^2 +16\left(9\beta\lambda+\nu\phi_2\right)\sigma^3\Big],
\end{equation}
\begin{multline}
g= \frac{1}{864}v^2\Big[\alpha ^6 \beta ^6 v^4+12 \alpha ^4 \beta ^5 \nu  v^4\sigma+ \left(12 \alpha ^4 \beta ^3 \phi_2 v^2+48 \alpha ^2 \beta ^4 \nu ^2 v^4\right)\sigma^2\\
+ \left(72 \alpha ^2 \beta ^2 (3 \beta  \lambda +\nu  \phi_2) v^2+64 \beta ^3 \nu ^3 v^4\right)\sigma^3\\
+\left(24 \alpha ^2 \phi_2^2+96 \beta  \nu  (9 \beta  \lambda +\nu  \phi_2) v^2\right)\sigma^4+864 \lambda  \sigma ^5 \phi_2\Big].
\end{multline}
The discriminant is proportional to $\sigma^6 v^4$, indicating that we
have an $\gsu(6)$ symmetry on $\sigma=0$ and an $\gsu(3)$ symmetry
tuned on $v=0$. The $\gsu(3)$ symmetry is the well-known non-Higgsable
cluster on $\mathbb{F}_3$, and since $[v]=S$, there is no matter
charged under both the $\gsu(6)$ and $\gsu(3)$ gauge groups, as
discussed above. We take the various parameters
(which are locally functions of the coordinates $u, v$) to have homology classes
\begin{align}
[\beta]=& 2F & [\alpha]=& S+ 3F & [\nu]=& 2F & [\lambda]=&0F & [\phi_2]=&0F.
\end{align}
Note that $\lambda$ and $\phi_2$ are essentially constants. 

We can now enhance the $\gsu(6)$ symmetry to $\gsu(7)$ by letting
\begin{align}
\beta=&\delta^2 & \alpha=&\delta\xi v & \phi_2=& 3 \kappa_0^2 & \lambda =& \rho_0 \kappa_0^3 & \nu =&\kappa_0\left(3\delta^2\rho_0 + \xi\right),
\end{align}
where
\begin{align}
[\delta]=& F & [\xi] =& 2F & [\kappa_0]=& 0F & [\rho_0]=& 0F.
\end{align}
The discriminant is now
\begin{equation}
\Delta = v^4\sigma^7\left[\frac{1}{8}v^6\delta^8\kappa_0^7\xi^4\left(6\rho_0\delta^2-\xi\right)+\mathcal{O}(\sigma)\right].
\end{equation}
To obtain an $\gsu(8)$ model, we therefore must let
\begin{equation}
\xi = 6\rho_0 \delta^2.
\end{equation}
This redefinition gives us an $\gsu(8)$ tuned on $\sigma=0$. In fact, we can set $\rho_0$ and $\kappa_0$ to $1$ without loss of generality, giving us a Weierstrass model of the form
\begin{multline}
y^2 = x^3 - 3 v^2\delta^2\left(9\delta^{18}v^6+18 \delta^{12}v^4\sigma + 15 \delta^6 v^2\sigma^2+4\sigma^3\right)x\\
+3v^2\left(18 \delta^{30}v^{10}+54 \delta^{24}v^{8}\sigma + 72 \delta^{18}v^6\sigma^2 + 48 \delta^{12}v^4\sigma^3 + 15\delta^{6}v^2\sigma^{4}+\sigma^{5}\right).
\end{multline}
The discriminant is now
\begin{equation}
\Delta = 27 v^4\sigma^8\left(9\delta^{12}v^4+14 \delta^{6}v^2\sigma + 9\sigma^2 \right).
\end{equation}
The $\sigma^8$ factor indicates that we have successfully tuned an $\gsu(8)$ gauge group, while the $v^4$ factor represents the expected non-Higgsable $\gsu(3)$ symmetry.

There are no hypermultiplets charged under the $\gsu(3)$ symmetry, but there are hypermultiplets charged under the $\gsu(8)$. Since $\sigma=0$ is a genus-two curve, there are two hypermultiplets in the adjoint ($\mathbf{63}$) representation of $\gsu(8)$. Additionally, the singularity type enhances from $I_8$ to $II^*$ at $\sigma=\delta=0$. Each of the two $\sigma=\delta=0$ points therefore supports hypermultiplets in the $\mathbf{56}+\mathbf{28}+\mathbf{8}$. In summary, the spectrum of hypermultiplets charged under the $\gsu(8)$ is
\begin{equation}
2\times\mathbf{63}+2\times\mathbf{56}+2\times\mathbf{28}+2\times\mathbf{8},
\end{equation}
which agrees with the $\gsu(8)$ anomaly cancellation conditions. If
this $\gsu(8)$ is Higgsed according to the Higgsing procedure of
\S\ref{sec:higgsprocess}, the resulting charge spectrum  becomes
\begin{multline}
14\times\chargehyper{8} +2\times\chargehyper{7} + 14\times\chargehyper{6} + 42\times\chargehyper{5}\\
+70\times\chargehyper{3} + 42\times\chargehyper{2} +14\times\chargehyper{1},
\end{multline}
satisfying the $\gu(1)$ anomaly conditions for $\uoneb{} = 8\times7\times2\tilde{S}$. Therefore, our explicit $\gsu(8)$ construction demonstrates that charge $\pmq{8}$ matter can be realized in F-theory. 

\subsection{Alternative Higgsings of SU(8) (charges 9, 10, 11, 12, 14)}
\label{sec:altHiggs}
In fact, one can obtain charges larger than $\pmq{8}$ through alternative Higgsings of this $\gsu(8)$ model. We will not perform an exhaustive investigation of all possible Higgsing chains, instead focusing on a few specific Higgsing processes that roughly follow the pattern
\begin{equation}
\gsu(8) \xrightarrow{\text{Higgs on }\mathbf{56}}\gsu(5)\times\gsu(3) \xrightarrow{\text{Higgs on }(\mathbf{24},\mathbf{1}),(\mathbf{1},\mathbf{8})} \gu(1)\times\gu(1) \xrightarrow{\text{Higgs on }(q_1,q_2)} \gu(1)
\end{equation}
Admittedly, this procedure is somewhat ad-hoc, suggesting that other Higgsing processes might produce even larger charges.

We start by giving \vev{}s to two full hypermultiplets of three-index
antisymmetric ($\mathbf{56}$) matter in the $\gsu(8)$ model considered
above.\footnote{D-term constraints for this breaking suggest we must
  give \vev{}s to two hypermultiplets instead of just one, just as for
  the Higgsing on two fundamental matter representations as described
  in \S\ref{sec:higgsprocess}.} This
\vev{} breaks the $\gsu(8)$ symmetry to $\gsu(5)\times\gsu(3)$, and
the $\gsu(8)$ representations branch as\footnote{Note that the
  branching patterns distinguish between representations and their
  conjugates. Even though full hypermultiplets still contain fields in
  $\mathbf{R}$ and $\overline{\mathbf{R}}$, it is important to keep
  track of representations and their conjugates for jointly charged
  matter. For instance, $(\mathbf{5},\overline{\mathbf{3}})$ is not
  the conjugate representation of $(\mathbf{5},\mathbf{3})$, and the
  two represent different types of hypermultiplets.}
\begin{align}
\mathbf{56} \rightarrow& \left(\overline{\mathbf{10}},\mathbf{1}\right) + \left(\mathbf{10},\mathbf{3}\right)+\left(\mathbf{5},\overline{\mathbf{3}}\right) + \left(\mathbf{1},\mathbf{1}\right)\\
\mathbf{63} \rightarrow& \left(\mathbf{24},\mathbf{1}\right) + \left(\mathbf{5},\overline{\mathbf{3}}\right) + \left(\overline{\mathbf{5}},\mathbf{3}\right)+\left(\mathbf{1},\mathbf{8}\right)+\left(\mathbf{1},\mathbf{1}\right)\\
\mathbf{28} \rightarrow& \left(\mathbf{10},\mathbf{1}\right) + \left(\mathbf{5},\mathbf{3}\right) + \left(\mathbf{1},\overline{\mathbf{3}}\right)\\
\mathbf{8}\rightarrow& \left(\mathbf{5},\mathbf{1}\right) + \left(\mathbf{1},\mathbf{3}\right). 
\end{align}
The two $(\mathbf{5},\overline{\mathbf{3}})$ hypermultiplets coming from the two $\mathbf{56}$ multiplets are eaten during the Higgsing process. Thus, after noting that hypermultiplets in conjugate representations are essentially the same, the $\gsu(5)\times\gsu(3)$ spectrum is
\begin{multline}
2\times\left(\mathbf{10},\mathbf{3}\right) + 4\times\left(\mathbf{5},\overline{\mathbf{3}}\right) +2\times\left(\mathbf{5},\mathbf{3}\right)\\
+ 2\times\left(\mathbf{24},\mathbf{1}\right)+2\times\left(\mathbf{1},\mathbf{8}\right) + 4\times\left(\mathbf{10},\mathbf{1}\right)+2\times\left(\mathbf{5},\mathbf{1}\right)+4\times\left(\mathbf{1},\mathbf{3}\right).
\end{multline}
As expected, this spectrum is consistent with the anomaly conditions
for $\gsu(5)$ and $\gsu(3)$ models tuned on two distinct divisors in
the homology class
$2\tilde{S}$. In fact, we construct an explicit F-theory realization of this $\gsu(5)\times\gsu(3)$ model in Appendix \ref{sec:su5su3weierstrass}.

Since there are two $\gsu(5)$ adjoint hypermultiplets  and two $\gsu(3)$ adjoint hypermultiplets, we can now Higgs the $\gsu(5)$ and $\gsu(3)$ groups individually using the Higgsing process in \S\ref{sec:higgsprocess}. The end result is a $\gu(1)\times\gu(1)$ gauge group. Hypermultiplets charged under this $\gu(1)\times\gu(1)$ symmetry are labeled as $(q_1,q_2)$, where $q_1$ and $q_2$ denote the charges under the two $\gu(1)$ symmetries. However, note that a $(q_1,q_2)$ hypermultiplet includes fields with charges $(q_1,q_2)$ and $(-q_1,-q_2)$. The branching patterns for the $\gsu(5)\times\gsu(3)$ representations are similar to the individual $\gsu(5)$ and $\gsu(3)$ branching described in \S\ref{sec:higgsprocess}. But it is important to note that charges coming from a conjugate representation $\overline{\mathbf{R}}$ are the negative of those coming from $\mathbf{R}$. To illustrate the effects of this fact, consider the branching patterns for the $(\mathbf{5},\mathbf{3})$ and $(\mathbf{5},\overline{\mathbf{3}})$ representations. First considering the $\gsu(5)$ and $\gsu(3)$ representations individually, the rules in \S\ref{sec:higgsprocess} suggest that the $\mathbf{5}$, $\mathbf{3}$, and $\overline{\mathbf{3}}$ representations branch as
\begin{align}
\mathbf{5}\rightarrow& 4\times\chargehyper{1} + \chargehyper{-4} \\ \mathbf{3}\rightarrow& 2\times\chargehyper{1} + \chargehyper{-2} \\ \overline{\mathbf{3}}\rightarrow& 2\times\chargehyper{-1} + \chargehyper{2}.
\end{align}
Note that we have kept track of the signs of the charges, and the signs for the charges coming from $\mathbf{3}$, and $\overline{\mathbf{3}}$ are negatives of each other. From these individual branching patterns, the $(\mathbf{5},\mathbf{3})$ representation should branch as
\begin{equation}
\left(\mathbf{5},\mathbf{3}\right)\rightarrow \left(-4,-2\right) + 2\times\left(-4,1\right) + 4\times\left(1,-2\right)+8\times\left(1,1\right).
\end{equation}
In contrast, the $(\mathbf{5},\overline{\mathbf{3}})$ representation should branch as
\begin{equation}
\left(\mathbf{5},\overline{\mathbf{3}}\right)\rightarrow \left(-4,2\right) + 2\times\left(-4,-1\right) + 4\times\left(1,2\right)+8\times\left(1,-1\right).
\end{equation}
These branching patterns are distinct from one another. For instance $\left(-4,-2\right)$ and $\left(-4,2\right)$ represent different types of multiplets, as the relative sign between the $q_1$ and $q_2$ charges differs. It is therefore important to distinguish between representations and their conjugates when considering the branching patterns.

In the end, the branching patterns for the $\gsu(5)\times\gsu(3)$ hypermultiplets are
\begin{align}
\left(\mathbf{10},\mathbf{3}\right)\rightarrow&4\times\left(-3,-2\right)+8\times\left(-3,1\right)+6\times\left(2,-2\right)+12\times\left(2,1\right)\\
\left(\mathbf{5},\mathbf{3}\right)\rightarrow& \left(-4,-2\right) + 2\times\left(-4,1\right) + 4\times\left(1,-2\right)+8\times\left(1,1\right)\\
\left(\mathbf{5},\overline{\mathbf{3}}\right)\rightarrow& \left(-4,2\right) + 2\times\left(-4,-1\right) + 4\times\left(1,2\right)+8\times\left(1,-1\right)\\
\left(\mathbf{24},\mathbf{1}\right)\rightarrow& 4\times\left(5,0\right)+4\times\left(-5,0\right)+16\times\left(0,0\right)\\
\left(\mathbf{1},\mathbf{8}\right)\rightarrow&2\times\left(0,3\right) + 2\times\left(0,-3\right) +4\times\left(0,0\right)\\
\left(\mathbf{10},\mathbf{1}\right) \rightarrow& 4\times\left(-3,0\right)+6\times\left(2,0\right)\\
\left(\mathbf{5},\mathbf{1}\right) \rightarrow& \left(-4,0\right)+4\times\left(1,0\right)\\
\left(\mathbf{1},\mathbf{3}\right) \rightarrow& \left(0,-2\right)+2\times\left(0,1\right).
\end{align}
To find the charged $\gu(1)\times\gu(1)$ spectrum, we must account for the fact that some of the $(\pm 5,0)$ and $(0,\pm 3)$ multiplets are eaten as part of the Higgsing process. Additionally, we are free to identify $(q_1,q_2)$ and $(-q_1,-q_2)$ hypermultiplets. Taking these facts into account, the charged $\gu(1)\times\gu(1)$ spectrum is
\begin{multline}
8\times\left(3,2\right)+16\times\left(3,-1\right)+12\times\left(2,-2\right)+24\times\left(2,1\right)\\
+2\times\left(4,2\right)+4\times\left(4,-1\right)+8\times\left(1,-2\right)+16\times\left(1,1\right)\\
+4\times\left(4,-2\right)+8\times\left(4,1\right)+16\times\left(1,2\right)+32\times\left(1,-1\right)\\
+8\times\left(5,0\right)+4\times\left(0,3\right)+16\times\left(3,0\right)+24\times\left(2,0\right)\\
+2\times\left(4,0\right)+8\times\left(1,0\right)+4\times\left(0,2\right)+8\times\left(0,1\right).
\end{multline}
This spectrum satisfies the $\gu(1)\times\gu(1)$ anomaly cancellation
conditions described in, for instance,
\cite{Erler6DAnom,ParkTaylorAbelian}.

Finally, we can Higgs $\gu(1)\times\gu(1)$ down to a single $\gu(1)$ by giving a \vev{} to a charged hypermultiplet. Suppose we give a \vev{} to a hypermultiplet with charge $(q_1^\prime, q_2^\prime)$. A hypermultiplet with $\gu(1)\times\gu(1)$ charge $(q_1,q_2)$ would then have a $\gu(1)$ charge given by
\begin{equation}
q = q_2^\prime q_1 - q_1^\prime q_2.
\end{equation}
Of course, the overall sign of $q$ is not too important, since
$\gu(1)$ charged hypermultiplets with charge $q$ have fields with
charges $+q$ and $-q$. Note that, at least for the charged $\gu(1)$
spectrum, we need not worry about the eaten degrees of freedom,
as they would have charge $q=0$. For particular $(q_1^\prime,
q_2^\prime)$, the resulting $\gu(1)$ charges can be higher than
$\pmq{8}$, as we illustrate with three examples:

\paragraph{Higgsing on charge $\mathbf{(1,2)}$ matter} If we give a \vev{} to $(q_1^\prime,q_2^\prime)=(1,2)$ matter, the resulting $\gu(1)$ charges are $2q_1-q_2$. Therefore, the $(4,-2)$ matter in the $\gu(1)\times\gu(1)$ spectrum would become charge $\pmq 10$ matter, while the $(4,-1)$ matter would become charge $\pmq 9$ matter. Indeed, the resulting charged $\gu(1)$ spectrum is
\begin{multline}
12\times\chargehyper{10}+4\times\chargehyper{9}+2\times\chargehyper{8}+24\times\chargehyper{7}+30\times\chargehyper{6}\\
+40\times\chargehyper{4}+60\times\chargehyper{3}+12\times\chargehyper{2}+24\times\chargehyper{1}.
\end{multline}
This spectrum satisfies the $\gu(1)$ anomaly conditions with $\uoneb = 86\times 2\tilde{S}$.
\paragraph{Higgsing on charge $\mathbf{(4,-1)}$ matter} Alternatively, if we Higgs on $(q_1^\prime,q_2^\prime)=(4,-1)$ matter, the resulting $\gu(1)$ charges are $q_1+4q_2$. Then, the $(0,3)$ matter becomes charge $\pmq 12$ matter, while the $(3,2)$ matter becomes charge $\pmq 11$ matter. The charged $\gu(1)$ spectrum is
\begin{multline}
6\times\chargehyper{12}+8\times\chargehyper{11}+16\times\chargehyper{9}+12\times\chargehyper{8}+8\times\chargehyper{7}\\
+36\times\chargehyper{6}+24\times\chargehyper{5}+14\times\chargehyper{4}\\
+48\times\chargehyper{3}+24\times\chargehyper{2}+24\times\chargehyper{1},
\end{multline}
which satisfies the $\gu(1)$ anomaly equations with $\uoneb=116\times2\tilde{S}$.

\paragraph{Higgsing on charge $\mathbf{(3,2)}$ matter} Finally, if we Higgs on $(q_1^\prime,q_2^\prime)=(3,2)$ matter, the resulting $\gu(1)$ charges are $2q_1-3q_2$. The $(4,-2)$ matter would becomes charge $\pmq 14$ matter. The charged $\gu(1)$ spectrum is
\begin{multline}
4\times\chargehyper{14} + 4\times\chargehyper{11}+20\times\chargehyper{10} + 20\times\chargehyper{9}+10\times\chargehyper{8}\\
+20\times\chargehyper{6}+40\times\chargehyper{5}+40\times\chargehyper{4}\\
+8\times\chargehyper{3}+10\times\chargehyper{2}+40\times\chargehyper{1},
\end{multline}
which satisfies the $\gu(1)$ anomaly cancellation conditions with $\uoneb=134\times2\tilde{S}$. 

To summarize, various Higgsing of the explicit $\gsu(8)$ F-theory
model above lead to charges $\pmq 9$, $\pmq 10$,$\pmq 11$,$\pmq 12$,
and $\pmq 14 $ (in addition to charges $\pmq 1$ through
$\pmq{8}$). Therefore, these charges should be realizable in
F-theory. Unluckily, the Higgsing processes considered here do not
produce charge $\pmq 13$ matter. But, given the ad-hoc nature of this
Higgsing process, it is likely that charge $\pmq 13$ can be realized
through some other means.

\subsection{SU(5)$\times$SU(4) (charges 15, 16, 20, 21)}
Even higher charges can be obtained by enhancing the $\gsu(5)\times\gsu(3)$ gauge group to $\gsu(5)\times\gsu(4)$. To obtain an explicit F-theory model realizing this $\gsu(5)\times\gsu(4)$ group, we take the $\gsu(5)\times\gsu(3)$ Weierstrass model described in Appendix \ref{sec:su5su3weierstrass} and set
\begin{equation}
\epsilon = 3\delta^2.
\end{equation}
The Weierstrass model is now
\begin{multline}
y^2=x^3 -3 v^2\delta^2\left(144 v^6 \delta^{18}-360 v^4\delta^{12}\sigma + 105 v^2\delta^{6}\sigma^2+10\sigma^3\right)x\\
+3 v^2\left(1152 v^{10}\delta^{30} - 4320 v^{8}\delta^{24}\sigma +3960v^{6}\delta^{18}\sigma^2-330v^{4}\delta^{12}\sigma^3+150v^{2}\delta^{6}\sigma^4+\sigma^5\right),
\end{multline}
where, as above, $[v]=S$,$[\delta]=F$, and $[\sigma]=2\tilde{S}$.
The discriminant meanwhile is
\begin{equation}
\Delta=27v^4\left(9\sigma-4v^2\delta^6\right)\left(\sigma-36 v^2\delta^6\right)^4\sigma^5,
\end{equation}
signaling an $\gsu(5)\times\gsu(4)$ gauge group with each factor tuned
on a divisor in the class $2\tilde{S}$. 

The only codimension-two locus with enhanced singularities is $\sigma=\delta=0$, where the singularity type enhances to $II^*$. By the Katz-Vafa analysis \cite{Katz:1996xe}, in which one breaks the $\mathbf{248}$ representation of $E_8$ to $\gsu(5)\times\gsu(4)$ representations, each of the two $\sigma=\delta=0$ points contributes
\begin{equation}
\left(\mathbf{10},\mathbf{1}\right)+\left(\mathbf{10},\mathbf{4}\right)+\left(\mathbf{5},\mathbf{6}\right)+\left(\mathbf{1},\mathbf{4}\right) + \left(\mathbf{5},\mathbf{\overline{4}}\right)
\end{equation}
hypermultiplets. Since both the $\gsu(5)$ and $\gsu(4)$ are tuned on genus-two curves, there are also two $(\mathbf{24},\mathbf{1})$ hypermultiplets and two $(\mathbf{1},\mathbf{15})$ hypermultiplets. Thus, the charged $\gsu(5)\times\gsu(4)$ spectrum is 
\begin{equation}
2\times\left[\left(\mathbf{10},\mathbf{1}\right)+\left(\mathbf{10},\mathbf{4}\right)+\left(\mathbf{5},\mathbf{6}\right)+\left(\mathbf{1},\mathbf{4}\right) + \left(\mathbf{5},\mathbf{\overline{4}}\right)+(\mathbf{24},\mathbf{1})+(\mathbf{1},\mathbf{15})\right].
\end{equation}
It is interesting to note that there is no anomaly-consistent model
with this gauge group and $b$ coefficients without the exotic
$\left(\mathbf{10},\mathbf{4}\right)$ matter.  Solving the anomaly
equations for generic matter fields and including only bifundamental
$\left(\mathbf{5},\mathbf{4}\right)$ fields would give rise to
negative multiplicities for some matter content.  Thus, this seems to
be the only Weierstrass model that realizes these gauge groups on
curves in the class $2 \tilde{S}$.

We can now give \vev{}s to the adjoint hypermultiplets as in \S\ref{sec:higgsprocess}, breaking $\gsu(5)\times\gsu(4)$ to $\gu(1)\times\gu(1)$. The $\gsu(5)\times\gsu(4)$ representations branch as follows:
\begin{align}
\left(\mathbf{10},\mathbf{4}\right)\rightarrow &4\times\left(-3,-3\right)+12\times\left(-3,1\right)+6\times\left(2,-3\right)+18\times\left(2,1\right), \\
\left(\mathbf{5},\mathbf{6}\right)\rightarrow & 3\times\left(-4,-2\right)+3\times\left(-4,2\right)+12\times\left(1,-2\right)+12\times\left(1,2\right),\\
\left(\mathbf{5},\mathbf{\overline{4}}\right)\rightarrow & 3\times\left(-4,-1\right)+1\times\left(-4,3\right)+12\times\left(1,-1\right)+4\times\left(1,3\right),\\
\left(\mathbf{10},\mathbf{1}\right)\rightarrow & 4\times\left(-3,0\right)+6\times\left(2,0\right),\\
\left(\mathbf{1},\mathbf{4}\right)\rightarrow & 1\times\left(0,-3\right)+3\times\left(0,1\right),\\
\left(\mathbf{24},\mathbf{1}\right)\rightarrow & 4\times\left(5,0\right)+16\times\left(0,0\right)+4\times\left(-5,0\right),\\
\left(\mathbf{1},\mathbf{15}\right)\rightarrow & 3\times\left(0,4\right)+9\times\left(0,0\right)+3\times\left(0,-4\right).
\end{align}
Accounting for the hypermultiplets eaten during the Higgsing process, the charged $\gu(1)\times\gu(1)$ spectrum is\footnote{Note that we have changed some of the signs by identifying hypermultiplets of charge $(-q_1,-q_2)$ with those of charge $(q_1,q_2)$.}
\begin{multline}
8\times\left(3,3\right)+24\times\left(3,-1\right)+12\times\left(-2,3\right)+36\times\left(2,1\right)\\
+6\times\left(4,2\right)+6\times\left(4,-2\right)+24\times\left(-1,2\right)+24\times\left(1,2\right)\\
+6\times\left(4,1\right)+2\times\left(4,-3\right)+24\times\left(-1,1\right)+8\times\left(1,3\right)\\
+8\times\left(3,0\right)+12\times\left(2,0\right)+2\times\left(0,3\right)+6\times\left(0,1\right)+8\times\left(5,0\right)+6\times\left(0,4\right).
\end{multline}

We now give a \vev{} to the charge $(4,-3)$ matter, which breaks $\gu(1)\times\gu(1)$ down to a single $\gu(1)$. The charges are given by $q=3q_1+4q_2$, leading to a charged spectrum of
\begin{multline}
8\times\chargehyper{21}+6\times\chargehyper{20}+12\times\chargehyper{16}+16\times\chargehyper{15}+2\times\chargehyper{12}\\
+24\times\chargehyper{11}+36\times\chargehyper{10}+8\times\chargehyper{9}+24\times\chargehyper{6}\\
+48\times\chargehyper{5}+12\times\chargehyper{4}+24\times\chargehyper{1}.
\end{multline}
The spectrum satisfies the $\gu(1)$ anomaly equations with
$\uoneb=372\times2\tilde{S}$. The spectrum includes charges as large
as $\pmq{15}$, $\pmq{16}$, $\pmq{20}$, and $\pmq{21}$. Moreover, the
charges above do not share any common factors, showing that the model
genuinely realizes these large charges. Therefore, the largest charge
that can be realized in F-theory must be at least charge $q
=\pmq{21}$.

\section{Open questions and further directions}
\label{sec:conclusion}
By combining explicit Weierstrass constructions of
6D supergravity theories having nonabelian gauge groups with the basic
physics of Higgsing processes, we have shown that F-theory can give rise to
charges as large as $q = \pm 21$ in 6D supergravity models with an abelian
U(1) gauge group.  We list here some questions  and open problems for
further research in this direction.
\vspace*{0.2in}

$\bullet$  While we have shown that charges up to $q = 21$ are
possible, we have not proven that this is the upper bound.  It would
be interesting to explore whether more exotic constructions can give
even higher U(1) charges in 6D F-theory models, and/or to prove an
upper bound on the charges allowed through F-theory constructions.

$\bullet$ We have noted that the constructions up to charge $q = 6$
follow from simpler F-theory models with less exotic singularity
types.  It would be nice to understand if there is a qualitative
difference between the geometric structure needed for charge $q \leq
6$ and that needed for charge $q > 6$.

$\bullet$ Since the abelian anomaly conditions allow for an infinite
set of solutions with U(1) gauge group and increasingly large charges,
even for models with no tensor multiplets ($T = 0$) \cite{AT-WT-infinite},
while only a finite number of F-theory models are possible, one may
look for new quantum consistency conditions on the low-energy theory
that may place an upper bound on the U(1) charge allowed in a
consistent theory.

$\bullet$ For nonabelian gauge groups such as SU($N$), the Kodaira
constraint from F-theory \cite{KMT-2} imposes a strict upper bound on the anomaly
coefficient $b$, namely $-12 a \geq N b$, meaning that $-12 a - N b$
lies in the positivity cone of the theory.  It is tempting to
speculate that there is a natural geometric constraint on the anomaly
coefficient $\tilde{b}$ for an abelian U(1) factor; the large size of
these coefficients for some of the constructions here, however
(e.g. $744 \tilde{S}$ for the model with abelian charges $q = \pm
21$), makes it clear that if there is such a bound it is quite large.
It would be nice to either prove the existence of such a bound from
geometry or give a convincing argument for the absence of such a bound.

$\bullet$ The constructions here are indirect and rely on the physical
mechanism of Higgsing.  To the best of our knowledge the largest $q$
that has been explicitly constructed in a Weierstrass model for a 6D
theory with only a U(1) gauge group is $q = 4$ \cite{Nikhil-34}.  It
would be good to have explicit constructions of the Weierstrass models
for higher abelian charges and to investigate the singularity
structure of the corresponding geometries.

$\bullet$ In particular, we have focused here on breaking SU($N$)
gauge groups to achieve high abelian charges.  It would be interesting
to explore whether breaking other groups such as exceptional groups
$E_7, E_6, F_4$ could also give large abelian charges.  Because the
constructions used here rely on exotic matter representations of
SU($N$), and such exotic matter representations cannot be realized in
a straightforward fashion for the exceptional groups \cite{KMRT}, it
may be harder to get large charges from other nonabelian groups;
nonetheless, this avenue should be explored more thoroughly.
(See also the following related point.)

$\bullet$ In the constructions in this paper we have relied on the
presence of nonabelian theories with exotic matter that has an
explicit construction through a Weierstrass model without
non-resolvable (4, 6) codimension 2 points that may be associated with
superconformal field theories.  It is possible that there may be
consistent abelian models with even higher charges that are related in
a similar way to ``unHiggsed'' nonabelian models with exotic matter
that gives rise to (4, 6) singularities in the geometry.  For example,
similar to the SU(5) $\times$ SU(4) model in the last section, one may
consider trying to construct a model with SU(6) $\times$ SU(3) gauge
group and matter in the $({\bf  15},{\bf  3})$ representation.
According to the logic of \cite{KMRT}, such a model would have a
singularity corresponding to an extended $\hat{E_8}$ Dynkin diagram,
which necessitates a (4, 6) point.  Even if this model is not
consistent as a nonabelian model, the Higgsed model with an abelian
factor may still be a valid F-theory construction.  There has also
been some recent suggestion that some exotic matter of this type may
give a consistent F-theory model as the singular features of the (4,
6) point can be compensated by T-brane degrees of freedom \cite{Cvetic:2018xaq}.
These questions would be interesting to understand further.

$\bullet$ The general structure of abelian charges when the gauge group
has both an abelian and nonabelian factor like $SU(N) \times U(1)$ has
been investigated in \cite{Lawrie:2015hia, Grimm:2015wda,
  CveticLinU1}.  It could be interesting to study the
range of charges that may be realized in explicit constructions with
such gauge groups.

$\bullet$ While the constructions here were carried out in 6
dimensions, where we have the strongest analytic control over F-theory
and the low-energy constraints are strongest, in principle the
Weierstrass constructions that lead to large $q$ charges should be
equally valid in four dimensions, although the story is complicated by
the presence of fluxes and the superpotential.  It would be
interesting to attempt explicit constructions of four-dimensional
F-theory models that give vacua with an abelian U(1) gauge theory and
similar large charges.

$\bullet$ The  abelian charges constructed here are much larger than
those realized in most other approaches to string compactification.
It would be interesting to systematically analyze other constructions
such as heterotic, type II and M theory on $G_2$ to see if a clear
upper bound on abelian charges can be demonstrated in those frameworks.

\acknowledgments

We would like to thank
Andres Collinucci, Craig Lawrie,
David Morrison, Paul Oehlmann, Andrew Turner and Roberto Valandro for helpful discussions.
This material is based upon work supported by the U.S.\ Department of
Energy, Office of Science, Office of High Energy Physics under
grant Contract Number
DE-SC00012567.
The work of NR is also supported by NSF grant PHY-1720321.
WT would like to thank the Aspen Center for Physics
for hospitality during the last stages of this work; the Aspen Center
for Physics is supported by National Science Foundation grant
PHY-1607611.


\appendix
\section{Explicit SU(5)$\times$SU(3) Weierstrass model}
\label{sec:su5su3weierstrass}
The $f$ and $g$ for the $\gsu(5)\times\gsu(3)$ Weierstrass model obtained by Higgsing the $\gsu(8)$ model above are
\begin{multline}
f=-3 v^2\Bigg[9 \delta^{12} v^6 \left(\delta^2-\epsilon\right)^4+18 \delta^8 \sigma  v^4 \left(\delta^2-2 \epsilon\right) \left(\delta^2-\epsilon\right)^2\\+3 \delta^4 \sigma ^2 v^2 \left(5 \delta^4-8 \delta^2 \epsilon+6 \epsilon^2\right)+2 \sigma ^3 \left(2 \delta^2+\epsilon\right)\Bigg]
\end{multline}
and
\begin{multline}
g=3 v^2 \Bigg[18 \delta^{18} v^{10} \left(\delta^2-\epsilon\right)^6+54 \delta^{14} \sigma  v^8 \left(\delta^2-2 \epsilon\right) \left(\delta^2-\epsilon\right)^4\\
+18 \delta^{10} \sigma ^2 v^6 \left(\delta^2-\epsilon\right)^2 \left(4 \delta^4-10 \delta^2 \epsilon+9 \epsilon^2\right)\\+6 \delta^6 \sigma ^3 v^4 \left(8 \delta^6-21 \delta^4 \epsilon+15 \delta^2 \epsilon^2-5 \epsilon^3\right)+15 \delta^2 \sigma ^4 v^2 \left(\delta^4+\epsilon^2\right)+\sigma ^5\Bigg].
\end{multline}
As before, the homology classes of $v$, $\sigma$ and $\delta$ are respectively $S$, $2\tilde{S}$ and $F$. The homology class of $\epsilon$ is $2F$. One recovers the original $\gsu(8)$ model when $\epsilon$ is taken to 0. In principle, one can obtain these expressions by tuning $f$ and $g$, as in \cite{AndersonGrayRaghuramTaylorMiT}: one expands $f$ and $g$ as series in $\sigma$ and and tunes the various parameters to force the discriminant to vanish to higher and higher orders. 

The discriminant is
\begin{equation}
\Delta=-27 v^4 \sigma ^5 \left(12 \delta^4 v^2 \epsilon-\sigma \right)^3 \left[9 \sigma ^2+9 v^4 \left(\delta^3-\delta \epsilon\right)^4+2 \sigma  v^2 \left(7 \delta^2-16 \epsilon\right) \left(\delta^2-\epsilon\right)^2\right].
\end{equation}
The $v^4$ factor in the discriminant reflects the expected $\gsu(3)$ non-Higgsable gauge group. Indeed, there are no codimension-two singularities along $v=0$ where the singularity type enhances, indicating that no matter is charged under the non-Higgsable $\gsu(3)$. However, the discriminant also has a $\sigma^5$ factor and a $(\sigma-12 \delta^4 v^2 \epsilon)^3$ factor. One can verify that $f$ and $g$ are not proportional to $\sigma$ or $(\sigma-12 \delta^4 v^2 \epsilon)$ and that the $\gsu(N)$ split conditions are satisfied.  These two factors in the discriminant therefore signal the presence of an $\gsu(5)$ group and an additional $\gsu(3)$ group, which together form the $\gsu(5)\times\gsu(3)$ symmetry coming from the Higgsed $\gsu(8)$. 

Since the $\gsu(5)$ and $\gsu(3)$ groups both occur on curves of genus
two, there are two hypermultiplets of $(\mathbf{24},\mathbf{1})$
matter and two hypermultiplets of $(\mathbf{1},\mathbf{8})$
matter. There are also four types of codimension-two loci that
contribute charged matter. On $\sigma=\epsilon-\delta^2=0$, the $I_5$
singularity type for the $\gsu(5)$ symmetry enhances to $I_1^*$. This
locus therefore contributes four hypermultiplets of
$(\mathbf{10},\mathbf{1})$ matter. On $\sigma-12 \delta^4 v^2
\epsilon=\epsilon-3\delta^2=0$, the $I_3$ singularity type for the
$\gsu(3)$ symmetry enhances to $I_4$, indicating that this locus
contributes four hypermultiplets of $(\mathbf{1},\mathbf{3})$
matter. Note that while the singularity type does enhance from $I_3$
to $II$ on $\sigma-12 \delta^4 v^2 \epsilon=5\epsilon+\delta^2=0$,
this locus does not contribute any charged matter.

The remaining codimension-two loci correspond to intersections between $\sigma=0$ and $\sigma-12 \delta^4 v^2 \epsilon=0$. Before describing the matter supported at these loci, it is worth mentioning a subtle point regarding the $\gsu(3)$. We can interpret the $I_3$ singularity on $\sigma-12 \delta^4 v^2 \epsilon=0$ as either an $\gsu(3)$ or an $\overline{\gsu(3)}$ gauge group. In other words, we can freely conjugate the $\gsu(3)$ symmetry, which will change the field-theoretic interpretation of the matter content. The $(\mathbf{1},\mathbf{3})$ hypermultiplets would be unaffected by this conjugation, as a single $(\mathbf{1},\mathbf{3})$ hypermultiplet contains fields in both the  $(\mathbf{1},\mathbf{3})$ and  $(\mathbf{1},\overline{\mathbf{3}})$ representations. But the jointly charged representations would be affected by this conjugation. For instance, a $(\mathbf{10},\overline{\mathbf{3}})$ hypermultiplet would become a  $(\mathbf{10}, {\mathbf{3}})$ after conjugation, and vice versa. Similarly, a $(\mathbf{5},\overline{\mathbf{3}})$ hypermultiplet would become a $(\mathbf{5},{\mathbf{3}})$ hypermultiplet.

With this in mind, we can analyze the loci where the two curves
intersect. At $\sigma=\epsilon=0$, the singularity type enhances to
$I_{8}$ or $A_{7}$. This locus therefore supports four hypermultiplets
bifundamental matter. But without performing an explicit resolution,
we cannot determine whether this locus supports four hypermultiplets
of $(\mathbf{5},\mathbf{3})$ matter or
$(\mathbf{5},\overline{\mathbf{3}})$ matter. Meanwhile, the
singularity type enhances to $III^*$ at $\delta=\epsilon=0$. The
Katz-Vafa method would suggest that, to determine the matter content,
we should break the adjoint ($\mathbf{133}$) representation of $E_7$
into $\gsu(5)\times\gsu(3)$ representations:
\begin{equation}
\mathbf{133} \rightarrow (\mathbf{24},\mathbf{1})+ (\mathbf{1},\mathbf{8}) + (\mathbf{10},\overline{\mathbf{3}}) + (\overline{\mathbf{10}},\mathbf{3}) + (\mathbf{5},\overline{\mathbf{3}}) + (\overline{\mathbf{5}},\mathbf{3}) +(\mathbf{5},\mathbf{1})+(\overline{\mathbf{5}},\mathbf{1})+2\times(\mathbf{1},\mathbf{1}).
\end{equation}
Naively, the $\delta=\epsilon=0$ locus would therefore seem to support two hypermultiplets of $(\mathbf{10},\overline{\mathbf{3}}) $ matter, two hypermultiplets of $(\mathbf{5},\overline{\mathbf{3}})$ matter, and two hypermultiplets of $(\mathbf{5},\mathbf{1})$ matter. However, the Higgsing patterns described in \S\ref{sec:altHiggs} suggests we should obtain $(\mathbf{10},\mathbf{3})$ matter, \emph{not} $(\mathbf{10},\overline{\mathbf{3}})$. To match the Katz-Vafa result with the field theory expectations, we should therefore conjugate the $\gsu(3)$. The $\delta=\epsilon=0$ locus then supports  $(\mathbf{10},{\mathbf{3}}) $ matter, two hypermultiplets of $(\mathbf{5},{\mathbf{3}})$ matter, and two hypermultiplets of $(\mathbf{5},\mathbf{1})$ matter, exactly as expected from the $\gsu(8)$ branching patterns. Moreover, the $\sigma=\epsilon=0$ should support four hypermultiplets of $(\mathbf{5},\overline{\mathbf{3}})$ matter. The complete charged spectrum for $\gsu(5)\times\gsu(3)$ is thus
\begin{multline}
2\times(\mathbf{10},\mathbf{3}) + 4\times(\mathbf{5},\overline{\mathbf{3}}) + 2\times(\mathbf{5},\mathbf{3})\\
+2\times(\mathbf{24},\mathbf{1})+2\times(\mathbf{1},\mathbf{8}) + 4\times(\mathbf{10},\mathbf{1}) + 2\times(\mathbf{5},\mathbf{1}) + 4\times(\mathbf{1},\mathbf{3}),
\end{multline}
which exactly matches expectations.

\section{Explicit Higgsing of SU(4)}
\label{sec:expHiggsingSU4}
While it is difficult to construct $\gu(1)$ models with large charges,
there are previous constructions admitting smaller charges. As a
result, the Higgsing process of section \ref{sec:higgsprocess} can be
seen explicitly in F-theory for small $\gsu(N)$. In this appendix, we
focus on the Higgsing of $\gsu(4)$ down to $\gu(1)$ with charge
$\pmq{4}$ matter. Explicit F-theory constructions with charge
$\pmq{4}$ matter were described in \cite{Nikhil-34}, and models admit
an unHiggsing that is the exact analogue of the $\gsu(4)\rightarrow
\gu(1)$ Higgsing process.

We start with an explicit charge-4 $\gu(1)$ model on an $\mathbb{F}_3$ base. The Weierstrass tuning (along with the section components, which we do not list here) was originally given in \cite{Nikhil-34}:
\begin{multline}
f=-\frac{1}{3} \left(s_5^2-3 s_1 s_8\right) \left(a_1^2 \left(d_1^2-3 d_0 d_2\right)-a_1 b_1 d_0 d_1+b_1^2 d_0^2\right)\\
-\frac{1}{3} \left(s_2^2-3 s_1 s_3\right) \left(a_1^2 d_2^2+b_1^2 \left(d_1^2-2 d_0 d_2\right)\right)\\
+\frac{1}{6} (2 s_2 s_5-3 s_1 s_6) \left(a_1^2 d_1 d_2+a_1 b_1 \left(d_1^2-2 d_0 d_2\right)+b_1^2 d_0 d_1\right)\\
+\frac{1}{6} (a_1 d_1+b_1 d_0) \left(2 b_1 d_2 \left(s_2^2-3 s_1 s_3\right)-3 s_2 s_6 s_8+s_5 \left(2 s_3 s_8+s_6^2\right)\right)\\
+a_1 d_0 \left(b_1 d_2 (3 s_1 s_6-2 s_2 s_5)+s_2 s_8^2-\frac{s_5 s_6 s_8}{2}\right)\\
+\frac{1}{6} (a_1 d_2+b_1 d_1) \left(s_3 (2 s_2 s_8-3 s_5 s_6)+s_2 s_6^2\right)\\
+\frac{1}{2} b_1 d_2 s_3 (2 s_3 s_5-s_2 s_6)-\frac{1}{48} \left(s_6^2-4 s_3 s_8\right)^2,
\end{multline}
\begin{multline}
g=\frac{1}{864} \left(s_6^2-4 s_3 s_8\right)^3-\frac{1}{2} \left(d_0 d_2^3 a_1^4+b_1^3 d_0 \left(d_1^3-3 d_0 d_1 d_2\right) a_1\right) s_1^2\\
+\frac{1}{4} \left(d_2^2 \left(d_1^2-2 d_0 d_2\right) a_1^4+b_1^2 \left(d_1^4-6 d_0^2 d_2^2-4 d_0 d_2 \left(d_1^2-2 d_0 d_2\right)\right) a_1^2+b_1^4 d_0^2 \left(d_1^2-2 d_0 d_2\right)\right) s_1^2\\
+\frac{1}{27} \left(\left(d_1^3-3 d_0 d_1 d_2\right) b_1^3+a_1^3 d_2^3\right) s_2 \left(9 s_1 s_3-2 s_2^2\right)\\
+\frac{1}{18} \left(d_1 d_2^2 a_1^3+b_1^2 \left(d_1^3-3 d_0 d_1 d_2\right) a_1+b_1^3 d_0 \left(d_1^2-2 d_0 d_2\right)\right) \left(\left(2 s_2^2-3 s_1 s_3\right) s_5-3 s_1 s_2 s_6\right)\\
+\frac{1}{18} \left(d_0 d_1 d_2 a_1^3+b_1 d_0 \left(d_1^2-2 d_0 d_2\right) a_1^2+b_1^2 d_0^2 d_1 a_1\right) \left(2 s_5^3-9 s_1 s_8 s_5+9 b_1 d_2 s_1^2\right)\\
\shoveleft{+\frac{1}{72} \left(d_1 d_2 a_1^2+b_1 \left(d_1^2-2 d_0 d_2\right) a_1+b_1^2 d_0 d_1\right)}\\ \times\Big[4 b_1 d_2 s_2 \left(2 s_2^2-9 s_1 s_3\right)+s_6 \left(s_6 (2 s_2 s_5+3 s_1 s_6)-12 s_3 s_5^2\right)\\
\shoveright{+4 \left(s_2 s_3 s_5-3 \left(s_2^2-5 s_1 s_3\right) s_6\right) s_8\Big]}\\
+\frac{1}{18} \left(d_2 \left(d_1^2-2 d_0 d_2\right) a_1^3+b_1 \left(d_1^3-3 d_0 d_1 d_2\right) a_1^2+b_1^3 d_0^2 d_1\right) \left(s_2 \left(2 s_5^2-3 s_1 s_8\right)-3 s_1 s_5 s_6\right)\\
+\frac{2}{9} \left(d_0 d_2^2 a_1^3+b_1^2 d_0 \left(d_1^2-2 d_0 d_2\right) a_1\right) \left(3 s_1 s_5 s_6+s_2 \left(3 s_1 s_8-2 s_5^2\right)\right)\\
+a_1^2 d_0^2 \left(-\frac{3}{2} b_1^2 d_2^2 s_1^2+\frac{1}{4} s_8^2 \left(s_5^2-4 s_1 s_8\right)+\frac{2}{9} b_1 d_2 s_5 \left(9 s_1 s_8-2 s_5^2\right)\right)\\
+\frac{1}{36} \left(\left(d_1^2-2 d_0 d_2\right) b_1^2+a_1^2 d_2^2\right) \left(3 \left(3 s_5^2-8 s_1 s_8\right) s_3^2+\left(4 s_2^2 s_8-3 s_6 (2 s_2 s_5+s_1 s_6)\right) s_3+2 s_2^2 s_6^2\right)\\
+\frac{1}{24} b_1 d_2 s_3 \left(6 b_1 d_2 s_3 \left(s_2^2-4 s_1 s_3\right)+(s_2 s_6-2 s_3 s_5) \left(s_6^2-4 s_3 s_8\right)\right)\\
\shoveleft{+\frac{1}{36} \left(d_0 d_2 a_1^2+b_1 d_0 d_1 a_1\right) \Big[\left(s_6^2+2 s_3 s_8\right) s_5^2+18 s_2 s_6 s_8 s_5-6 \left(s_2^2+2 s_1 s_3\right) s_8^2}\\
\shoveright{+4 b_1 d_2 \left(\left(2 s_2^2-3 s_1 s_3\right) s_5-3 s_1 s_2 s_6\right)-33 s_1 s_6^2 s_8\Big]}\displaybreak\\
-\frac{1}{54} \left(\left(d_1^3-3 d_0 d_1 d_2\right) a_1^3+b_1^3 d_0^3\right) \left(4 s_5^3+9 s_1 (3 b_1 d_2 s_1-2 s_5 s_8)\right)\\
\shoveleft{+\frac{1}{72} a_1 d_0 \Big[16 b_1^2 s_2 \left(9 s_1 s_3-2 s_2^2\right) d_2^2}\\
+6 b_1 \left(s_6 \left(6 s_3 s_5^2+s_6 (9 s_1 s_6-8 s_2 s_5)\right)+2 \left(3 \left(s_2^2+2 s_1 s_3\right) s_6-8 s_2 s_3 s_5\right) s_8\right) d_2\\
\shoveright{+3 s_8 (s_5 s_6-2 s_2 s_8) \left(s_6^2-4 s_3 s_8\right)\Big]}\\
+\frac{1}{18} \left(d_0 d_1 a_1^2+b_1 d_0^2 a_1\right) \left(2 b_1 d_2 \left(s_2 \left(2 s_5^2-3 s_1 s_8\right)-3 s_1 s_5 s_6\right)-3 s_8 \left(s_6 s_5^2+(s_2 s_5-6 s_1 s_6) s_8\right)\right)\\
-\frac{1}{72} (b_1 d_1+a_1 d_2) \left(12 b_1 d_2 s_3 \left(s_2 s_3 s_5+\left(s_2^2-6 s_1 s_3\right) s_6\right)+\left(s_6^2-4 s_3 s_8\right) \left(s_2 s_6^2+s_3 (2 s_2 s_8-3 s_5 s_6)\right)\right)\\
\shoveleft{+\frac{1}{72} (b_1 d_0+a_1 d_1) \Big[2 b_1 d_2 \left(-6 \left(s_5^2+2 s_1 s_8\right) s_3^2+\left(2 s_8 s_2^2+18 s_5 s_6 s_2-33 s_1 s_6^2\right) s_3+s_2^2 s_6^2\right)}\\
\shoveright{-\left(s_6^2-4 s_3 s_8\right) \left(s_5 \left(s_6^2+2 s_3 s_8\right)-3 s_2 s_6 s_8\right)\Big]}\\
\shoveleft{+\frac{1}{36} \left(\left(d_1^2-2 d_0 d_2\right) a_1^2+b_1^2 d_0^2\right) \Big[2 \left(s_6^2+2 s_3 s_8\right) s_5^2-6 s_2 s_6 s_8 s_5}\\
+8 b_1 d_2 \left(-2 s_5 s_2^2+3 s_1 s_6 s_2+3 s_1 s_3 s_5\right)-3 s_8 \left(s_1 \left(s_6^2+8 s_3 s_8\right)-3 s_2^2 s_8\right)\Big].
\end{multline}
We take the parameters to have the homology classes given in
Table
\ref{tab:q4homology}. Note that $d_0$ and $s_1$ have trivial homology
classes and are thus constants. Additionally, the homology classes
suggest that many of the parameters are reducible. While not
necessary, one can explicitly address this by setting
\begin{align}
d_1 =& v \tilde{d}_1 & d_2 =& v^2\tilde{d}_2 & s_5 =& v\tilde{s}_5 & s_6 =& \tilde{s}_6 & s_8 =&v^2\tilde{s}_8,
\end{align}
where $[v]=S$. Performing these redefinitions makes the $\gsu(3)$ non-Higgsable cluster (NHC) on $\mathbb{F}_3$ explicitly visible in the Weierstrass model. There are no nonabelian gauge groups other than this NHC, and there is no matter charged under the nonHiggsable $\gsu(3)$. Based on the matter analysis from \S4.3 of \cite{Nikhil-34}, the charged matter spectrum of the model is
\begin{equation}
6\times\chargehyper{4}+32\times\chargehyper{3}+60\times\chargehyper{2}+96\times\chargehyper{1}.
\end{equation}

\begin{table}
\begin{center}
\begin{tabular}{|cc|}\hline
Parameter & Homology Class\\\hline
$a_1$ & $3F$\\
$b_1$ & $2\tilde{S}=2S + 6F$\\
$d_0$ & $0S+0F$\\
$d_1$ & $2S + 3F$\\
$d_2$ & $4S+6F$\\
$s_1$ & $0S+0F$\\
$s_2$ & $F$\\
$s_3$ & $2F$\\
$s_5$ & $2S+4F$\\
$s_6$ & $2S+5F$\\
$s_8$ & $4S+8F$\\\hline
\end{tabular}
\end{center}
\caption{Homology classes for the parameters of the charge-4
  Weierstrass model on $\mathbb{F}_3$.}
\label{tab:q4homology}
\end{table}

We can then unHiggs the $\gu(1)$ symmetry to $\gsu(4)$ by setting $a_1\rightarrow 0$, $s_2\rightarrow0$, $s_3\rightarrow0$. One can verify that these tunings cause the generating section for the $\gu(1)$ to coincide with the zero section, suggesting that the $\gu(1)$ has been unHiggsed to some nonabelian gauge group. The discriminant now takes the form
\begin{equation}
\Delta = -\frac{1}{16}b_1^4 v^4 d_0^3 s_1^2\left[\tilde{s}_6^4\left(\tilde{d}_2\tilde{s}_6^2-\tilde{d}_1\tilde{s}_6\tilde{s}_8+d_0\tilde{s}_8^2\right)v^4+\mathcal{O}(b_1)\right].
\end{equation}
Because $s_1$ and $d_0$ are constants, the $d_0^3$ and $s_1^2$ factors in the discriminant do not represent any new nonabelian gauge groups. Meanwhile, the $v^4$ factors corresponds to the expect $\gsu(3)$ NHC on $\mathbb{F}_3$. But the $b_1^4$ factor represents a new $\gsu(4)$ gauge group.\footnote{$f$ and $g$ are not proportional to $b_1$ after the tunings, and the split condition is satisfied.} The $\gu(1)$ has therefore been unHiggsed to an $\gsu(4)$ tuned on $b_1=0$. Since $[b_1]=2\tilde{S}$, $b_1=0$ is a genus-two curve, and the spectrum includes two hypermultiplets in the adjoint $(\mathbf{15})$ representation. Additionally, the codimension-two locus $\tilde{s}_6=b_1=0$ contributes ten $\mathbf{10}$ hypermultiplets, while the locus $(\tilde{d}_2\tilde{s}_6^2-\tilde{d}_1\tilde{s}_6\tilde{s}_8+d_0\tilde{s}_8^2)=b_1=0$ contributes thirty-two $\mathbf{4}$ hypermultiplets. This charged spectrum agrees exactly with anomaly cancellation. 

We therefore see that the F-theory charge-4 model admits a $\gu(1)\rightarrow\gsu(4)$ unHiggsing. Therefore, the corresponding $\gsu(4)\rightarrow\gu(1)$ unHiggsing also occurs in F-theory, providing further evidence that the $\gsu(N)\rightarrow \gu(1)$ Higgsing of \S\ref{sec:higgsprocess} should be valid more generally.
\bibliographystyle{JHEP}
\bibliography{references}








\end{document}